\documentclass[12pt,nofootinbib,tightenlines,preprintnumbers,superscriptaddress]{revtex4}
\usepackage[T1]{fontenc}
\usepackage[utf8]{inputenc}
\usepackage{lmodern}
\usepackage[english]{babel}
\usepackage{graphicx}
\usepackage{amsmath,mathtools,amssymb}
\usepackage{hyperref}
\usepackage{tabularx}
\usepackage{hhline}
\usepackage{booktabs}
\usepackage[dvipsnames]{xcolor}
\usepackage{tikz}
\usepackage{bm}
\usepackage[normalem]{ulem}
\usetikzlibrary{decorations.pathreplacing,decorations.pathmorphing}

\usepackage{siunitx}
\usepackage{soul}

\usepackage{physics}

\DeclareFontFamily{U}{mathx}{\hyphenchar\font45}
\DeclareFontShape{U}{mathx}{m}{n}{<-> mathx10}{}
\DeclareSymbolFont{mathx}{U}{mathx}{m}{n}
\DeclareMathAccent{\widebar}{0}{mathx}{"73}

\newcommand{\beq}{\begin{equation}}
\newcommand{\eeq}{\end{equation}}

\newcommand{\sd}{\mathrm d}

\newcommand{\sT}{\mathrm T}

\newcommand{\sgn}{\mathrm{sgn}}
\newcommand{\eexp}{\mathrm{e}}
\newcommand{\I}{\mathrm{i}}

\newcommand{\om}{\omega}

\newcommand{\circp}{ p^{^{\!\!\!\circ}} }
\newcommand{\hatp}{\hat{p}}

\newcommand{\circq}{ q^{^{\!\!\!\circ}} }

\begin{document}

\preprint{MITP-23-053}
\title{Probing the photon emissivity of the quark-gluon plasma \\ without an inverse problem in lattice QCD }
\author{Marco C{\`e}}
\affiliation{Dipartimento di Fisica, Università di Milano-Bicocca, Piazza della Scienza 3, 20126 Milano, Italy}
\affiliation{INFN, Sezione di Milano-Bicocca, Piazza della Scienza 3, 20126 Milano, Italy}
\affiliation{Albert Einstein Center for Fundamental Physics
and Institut für Theoretische Physik,
Universität Bern, Sidlerstrasse 5, 3012 Bern, Switzerland}

\author{Tim Harris}
\affiliation{Institute for Theoretical Physics, ETH Zürich, Wolfgang-Pauli-Str. 27, 8093 Zürich, Switzerland}

\author{Ardit Krasniqi}
\affiliation{$PRISMA^+$ Cluster of Excellence \& Institut für Kernphysik, Johannes Gutenberg-Universität Mainz,\\
Saarstr. 21, 55122 Mainz, Germany}

\author{Harvey B. Meyer}
\affiliation{$PRISMA^+$ Cluster of Excellence \& Institut für Kernphysik, Johannes Gutenberg-Universität Mainz,\\
Saarstr. 21, 55122 Mainz, Germany}
\affiliation{Helmholtz Institut Mainz, Johannes Gutenberg-Universität Mainz,
Saarstr. 21, 55122 Mainz, Germany}

\author{Csaba T\"or\"ok}
\affiliation{$PRISMA^+$ Cluster of Excellence \& Institut für Kernphysik, Johannes Gutenberg-Universität Mainz,\\
Saarstr. 21, 55122 Mainz, Germany}


\begin{abstract}
{

The thermal photon emissivity of the quark-gluon plasma is determined
by the in-medium spectral function of the electromagnetic current at
lightlike kinematics, $\sigma(\omega)$.  In this work, we present the
first lattice QCD results on moments of $\sigma(\omega)/\omega$,
defined by the weight function $1/(\omega^2+ (2\pi T n)^2)$, $n\in\mathbb{Z}$
and computed without encountering an inverse problem.
We employ two dynamical flavours of O($a$)-improved Wilson fermions at a temperature $T\approx 250\;$MeV
and perform the continuum limit.
We compare our results for the first two moments to
those obtained dispersively by integrating over the spectral function
computed at weak coupling by Arnold, Moore and Yaffe.
}
\end{abstract}

\maketitle

\section{Introduction}
\label{sec:intro}

The strongly-interacting matter created during a heavy-ion collision radiates
electromagnetic energy due to the interactions of its charged constituents.
Photons, the quanta of this radiation, have a large mean free path compared to
the strongly-interacting degrees of freedom, thanks to the hierarchy of the
electromagnetic and strong coupling constants at experimentally accessible 
temperatures, $\alpha_\mathrm{em}\simeq 1/137 \ll \alpha_s$.
These probes escape from the strongly interacting medium and can inform us
about its entire space-time evolution.

Let $\omega$ be the energy of a photon emitted from a fluid cell at rest and in thermal equilibrium.
The differential rate of photon emission per unit volume of the cell
is described to all orders in the strong coupling, and to
leading order in the electromagnetic coupling by the formula~\cite{McLerran:1984ay}
\begin{align}
    \frac{\mathrm d\Gamma_\gamma(\omega)}{\mathrm d\omega} &=
        \frac{\alpha_\mathrm{em}}{\pi}
        \frac{2\omega\sigma(\omega)}{\mathrm e^{\omega/T}-1}
        + \mathrm O(\alpha_\mathrm{em}^2),
    \label{eq:photon_rate}
\end{align}
where $\sigma(\omega)$ is the in-medium spectral function associated with the two-point correlation function
of the electromagnetic current at zero-virtuality. 
Therefore, this temperature-dependent spectral function directly determines the
emissivity of real photons from thermal matter described by quantum chromodynamics (QCD).
From the experimentally measured spectrum of photons,
one would hope to infer interesting physical properties of the thermal medium, starting with its temperature.
However, since the QCD matter created in a heavy-ion collision does not remain in
equilibrium, the prediction of the total yield is complicated by
the necessity to integrate over the spacetime history of the expanding fluid
as well as to include non-thermal components~\cite{David:2019wpt}.

Nevertheless, recent experimental results from the RHIC and the LHC facilities
have been compared to phenomenological models which incorporate a thermal component
produced by the expanding plasma~\cite{Gale:2021zlc}.
Data for the yield of direct photons from the PHENIX and ALICE experiments at
RHIC and the LHC, respectively, show an excess of production in the region of phase space where
thermal sources dominate, i.e.\ for transverse momenta, $p_\mathrm{T}$, below a few
$\mathrm{GeV}$~\cite{PHENIX:2022rsx,PHENIX:2014nkk,ALICE:2015xmh,Gale:2021zlc}.
In contrast, data from the STAR experiment at RHIC appear to be in agreement with the
available theoretical models~\cite{STAR:2016use}.
Furthermore, the anisotropy of the photon momentum distribution, the photon
$v_2$, is larger for all experiments than models suggest~\cite{PHENIX:2011oxq,PHENIX:2015igl,ALICE:2018dti}.

The failure of most models in describing collectively the  experimental findings mentioned above
is usually referred to as the direct photon 
puzzle~\cite{David:2019wpt,Geurts:2022xmk,Gale:2021zlc}.
The model calculations rely on theoretical predictions of the 
thermal photon emissivity, which is directly proportional to $\sigma(\omega)$.
So far, the phenomenological models have used such predictions obtained at leading
order in QCD perturbation theory~\cite{Arnold:2001ms,Arnold:2001ba}, 
complemented with relativistic kinetic theory calculations in a hot meson gas 
for the hadronic phase (see e.g.~\cite{Paquet:2015lta,Turbide:2003si}).
How exactly to connect these predictions in the vicinity of the
crossover temperature of about 150\;MeV is unclear, hence interpolations
have been used.

Lattice QCD is a suitable framework for the non-perturbative determination
of the correlation function which mathematically determines $\sigma(\omega)$ uniquely.
However, given that lattice QCD is formulated in Euclidean space,
obtaining the spectral density by analytic continuation from the current
correlator is a numerically ill-posed problem (see for instance the reviews~\cite{Meyer:2011gj,Aarts:2020dda}).
The analytic continuation needed to determine real-time quantities
from the correlators computed in Euclidean spacetime is further hampered by the
fact that a relatively small number of noisy data points are available,
usually $\mathcal{O}(10-50)$ at finite temperature.
Yet what one is interested in is determining a spectral function at
$\mathcal{O}(1000)$ values of $\omega$ covering the interesting kinematical
region with a sufficiently high resolution and with a few-percent precision.

Aiming at a goal which is within reach of the current numerical capabilities,
one can try to extract not the spectral function itself, but a smeared version thereof.
In the method originally proposed by G.\ E.\ Backus and J.\ F.\ Gilbert~\cite{Backus:1968abc,Backus:1970},
the calculation of a smeared, filtered spectral function is viable after
determining certain coefficients that multiply the Euclidean correlator data points.
The value at a given $\om$ of this smeared spectral function is a weighted
average of the actual spectral function over the vicinity of $\omega$.
The weighted average is defined as an integral over the spectral function multiplied by a compact
kernel centered at a certain value of $\om$.
This method has been applied to several
problems~\cite{Brandt:2015aqk,Hansen:2017mnd,Ce:2022dax,Ce:2022fot,Altenkort:2022yhb,Astrakhantsev:2019zkr,Astrakhantsev:2018oue,Alexandrou:2022tyn}
in lattice QCD, and similar approaches have been proposed which offer a path to reducing
the width of the filtering kernel, thus getting closer to the unsmeared spectral 
function~\cite{Hansen:2019idp,Bulava:2021fre,Gambino:2022dvu}.
These approaches have in common that they
provide information about an integral over the spectral function
based on the Euclidean correlator and additional technical ingredients.

In this work, we obtain integrals over $\sigma(\omega)$ by calculating a correlator directly accessible in lattice QCD
that has be shown~\cite{Meyer:2018xpt} to have a simple integral representation in terms of $\sigma(\omega)$.
The key ingredients are to write a dispersion relation for the Euclidean correlator at fixed photon virtuality,
rather than at fixed spatial momentum, and to employ an imaginary spatial momentum in order to realize lightlike kinematics
in Euclidean space. We thus obtain information about the spectral function at lightlike kinematics
without facing an inverse problem.
This cannot be achieved when employing dispersion relations at fixed spatial photon momentum, which has been the only type
of dispersion relation applied to date in lattice-QCD based studies of non-equilibrium properties
of the quark-gluon plasma, starting with Ref.~\cite{Karsch:1986cq}.

Specifically, we compute the first two moments of $\sigma(\omega)/\omega$,
defined by the weight function $1/(\omega^2+ (2\pi T n)^2)$ for $n=1$ and $n=2$.
The construction makes use of the spatially transverse Euclidean correlator
evaluated at Matsubara frequency $\om_n$ and at imaginary spatial momentum,
$k=\I \om_n$. Such an `imaginary momentum' corresponds to a weight function
$\exp(\om_n x_3)$ in the spatial coordinate $x_3$, which strongly enhances the contribution from large positive $x_3$.
Thus the control over the statistical fluctuations at long distances is the main difficulty in computing these observables in lattice QCD.
Numerically, the task thus bears a strong resemblance with the lattice calculation of the hadronic vacuum polarization
contribution to the muon $g-2$~\cite{Meyer:2018til,Aoyama:2020ynm,Borsanyi:2020mff}.
Using the spectral function $\sigma(\omega)$ computed in QCD at weak coupling~\cite{Arnold:2001ms,Arnold:2001ba},
as well as that computed in strongly coupled super-Yang-Mills by AdS/CFT methods~\cite{CaronHuot:2006te},
we compute the same moments and compare them to our results.
It is the first time that the weak-coupling QCD predictions for the photon emissivity of the quark-gluon plasma
can be tested non-perturbatively without any uncertainties associated with an ill-posed inverse problem.

The structure of the paper is the following:
In Sec.~\ref{sec:cont.defs.}, we introduce the basic observables in
the continuum.
In Sec.~\ref{sec:subtractions}, we discuss various subtractions that are
necessary to reduce large cutoff effects afflicting our primary observable,
while Sec.~\ref{sec:dHEdQ2} discusses the virtuality dependence of this
observable.
The lattice observables are defined in Sec.~\ref{sec:lat.defs.}.
Thereafter, in Sec.~\ref{sec:simul}, we provide some details about our numerical setup.
This is followed by the presentation of the main results in Sec.~\ref{sec:results} and Sec.~\ref{sec:res,HE2},
in which we devote different subsections to the different derived observables
that we investigated and describe our analysis for the sectors $n=1$ and $n=2$, respectively.
In Sec.~\ref{sec:comp}, we compare our results to predictions made
using either free quarks, the weak-coupling expansion or the strongly-coupled
$\mathcal{N}=4$ super Yang-Mills theory (SYM).
Finally, we conclude in Sec.~\ref{sec:conclusions}.

\section{Preliminaries}
\subsection{Definitions in the continuum}
\label{sec:cont.defs.}
The spectral function of the electromagnetic current is defined as
\beq
	\rho_{\mu\nu}(\om,{\bf k}) = \int \sd^4 x\, e^{\I(\om t - {\bf k}{\bf x})}\, 
	\langle [J_\mu^{\rm{em}}(x), J_\nu^{\rm{em}}(0)^\dagger] \rangle.
\eeq
Here, $J_\mu^{\rm{em}}(x) = \sum_f Q_f \bar{\psi}_f(x) \gamma_\mu \psi_f(x)$
is the electromagnetic current, $Q_f\in\{2/3,-1/3\}$ denotes the charge of a quark with flavor $f$
and $\gamma_{\mu}$ are the Euclidean Dirac matrices satisfying
$\{\gamma_\mu,\gamma_\nu\} = 2\delta_{\mu\nu}$, i.e.\ ${\gamma_\mu}^\dagger=\gamma_\mu$.
The time evolution is given in real time by $J_\mu^{\rm{em}}(x)= e^{\I H t} J_\mu^{\rm{em}}(0) e^{-\I H t}$.
The expectation value is taken with respect to the thermal density matrix $\eexp^{-\beta H}/Z$,
where $Z$ is the grand canonical partition function, $H$ is the QCD Hamiltonian and $\beta=1/T$
is the inverse temperature.
The thermal photon production rate per unit volume of the QGP, $\sd \Gamma_\gamma(\om)/{\sd \om}$,
is related to the transverse channel spectral function evaluated at
lightlike kinematics, $\sigma(\om) \equiv \rho_\sT(\om, k=\om)$, where 
\beq
\rho_\sT(\om, k) = \frac{1}{2}(\delta^{ij} - k^i k^j/k^2) \rho_{ij}(\om,{\bf k})
\eeq
is the transverse channel spectral function.

We remind the reader that Eq.~\eqref{eq:photon_rate} is valid at leading order 
in the electromagnetic coupling constant, but to all orders in the strong 
coupling constant~\cite{McLerran:1984ay}.

Let us consider now the Euclidean screening correlators\footnote{From here on, we drop the superscript `em' on the vector current, since the following equations are independent of its specific flavour structure.}
\beq
G_{E,\,\mu\nu}(\om_n, p_1, p_2; x_3) = - \int_0^{\beta} \sd x_0 \,\eexp^{\I \om_n x_0} \int \sd x_1 \sd x_2 \,\eexp^{\I (p_1 x_1 + p_2 x_2)} \,\langle J_{\mu}(x_0,x_1,x_2,x_3) J_{\nu}(0) \rangle,
	\label{eq:scr_corr,gen}
\eeq
with $\om_n= 2 n \pi T$ being the $n$th Matsubara frequency and the time-evolution
being given in Euclidean space-time ($O(x_0)= \eexp^{H x_0}\, O\, \eexp^{-H x_0}$).
In the presence of interactions, the low-lying spectrum of the spectral function 
corresponding to Eq.~(\ref{eq:scr_corr,gen}) is gapped and discrete, resulting in an exponential 
falloff for $G_{E,\,\mu\nu}(\om_n, p_1, p_2; x_3)$.
The corresponding energies are called screening masses and have been investigated
in Ref.~\cite{Brandt:2014uda} using weak-coupling theory as well as lattice simulations.
We discuss these in Sec.~\ref{sec:E0,cont} and in App.~\ref{app:weak-cth}.

Restricting the further discussion to the transverse channel, we define
\beq
	G_{E}^{\sT}(\om_n, p; x_3) \equiv G_{E, 11}(\om_n, 0, p; x_3) = - \int_0^\beta \sd x_0\, \eexp^{\I \om_n x_0} \int \sd x_1 \sd x_2 \,\eexp^{\I p x_2} \,\langle J_1(x) J_1(0) \rangle
	\label{eq:GT,cont}
\eeq
and two special cases of Eq.~(\ref{eq:GT,cont}), the non-static (ns) and static (st) screening 
correlators in the transverse channel as
\beq
	G_{\rm{ns}}^\sT(\om_n, x_3) \equiv G_{E}^{\sT}(\om_n, p=0; x_3),
	\label{eq:GT,ns,cont}
\eeq
and
\beq
	G_{\rm{st}}^\sT(p, x_3) \equiv G_{E}^{\sT}(\om_n=0, p; x_3),
	\label{eq:GT,st,cont}
\eeq
respectively.
Thus, for the non-static (static) screening correlator, the momentum is inserted 
into the temporal (spatial) direction.

The Fourier transform of the non-static screening correlator is defined as
\beq
	\tilde G_{\rm{ns}}^{\sT}(\om_n, k) = \int_{-\infty}^{\infty} \sd x_3\, G_{\rm{ns}}^\sT(\om_n, x_3)\, \eexp^{\I k x_3}.
	\label{eq:GT,ns,Fourier}
\eeq
Evaluating Eq.~(\ref{eq:GT,ns,Fourier}) at imaginary spatial momentum, we define~\cite{Meyer:2018xpt}
\beq
	H_E(\om_n) \equiv \tilde G_{\rm{ns}}^\sT(\om_n, k=\I \om_n),
\eeq
the spatially transverse Euclidean correlator evaluated at Matsubara frequency $\om_n$
and at imaginary spatial momentum $k=\I\om_n$.
Correspondingly, we can also write
\begin{align}
	H_E(\om_n) &= \int_{-\infty}^{\infty} \sd x_3\, G_E^\sT(\om_n, 0; x_3)\, \eexp^{- \om_n x_3} \\
			   &= -\int_0^\beta \sd x_0 \int \sd^3 x\,\, \eexp^{\I \om_n x_0}\, \eexp^{-\om_n x_3}\, \langle J_1(x) J_1(0) \rangle.
   \label{eq:HE,intGscr}
\end{align}
An important property of $H_E(\om_n)$ is that it vanishes identically in the vacuum,
a consequence of Lorentz symmetry~\cite{Meyer:2018xpt}.
Secondly, subtracting explicitly $H_E(0)$, which vanishes due to current conservation,
suffices to make the expression on the right-hand side of Eq.\ (\ref{eq:HE,intGscr}) finite by power counting.
Indeed, this subtraction amounts to the modification $e^{\om_n(ix_0 - x_3)}\to (e^{\om_n(ix_0 - x_3)} -1)$ inside the integral,
and Taylor-expanding the exponential function yields the leading term $\frac{1}{2}\om_n^2 (x_3^2 -x_0^2)$
if we drop terms that do not contribute due to the $x_3\to-x_3$ or $x_0\to -x_0$ symmetry of the $\langle J_1(x) J_1(0) \rangle$ correlator.
The terms $\frac{1}{2}\om_n^2 (x_3^2 -x_0^2)$, which separately would generate a logarithmic UV-divergence, obviously cancel each other
exactly in the vacuum, and therefore only make a UV-finite contribution at non-zero temperature.
This last statement is easily proven by power counting, using the operator-product expansion.
Even though the simple subtraction $(e^{\om_n(ix_0 - x_3)} -1)$ is thus in principle sufficient,
in subsection \ref{sec:subtractions} we will derive other possible subtractions,
that turn out to be superior in practical lattice QCD calculations,
and test them  in the context of lattice perturbation theory (appendix~\ref{app:lattpt}),

In Ref.~\cite{Meyer:2018xpt} it has been shown that the spectral function
at vanishing virtuality, $\sigma(\om)$, is related to $H_E(\om_n)$
via the following once-subtracted dispersion relation
\beq
	H_E(\om_n) - H_E(\om_r) = \int_0^\infty \frac{\sd \om}{\pi}\, \om \sigma(\om) \left[ \frac{1}{\om^2+\om_n^2} - \frac{1}{\om^2+\om_r^2} \right].
	\label{eq:HEn-HEr,disprel}
\eeq
The left-hand side contains the difference of two observables
that can be evaluated in Euclidean space-time and this difference
directly probes an integral of the spectral function at vanishing virtuality.

Given that $H_E(0)=0$ in the absence of massless static screening modes,
a simplified dispersion relation applies for $H_E(\om_n)$,
\beq
	H_E(\om_n) = -\frac{\om_n^2}{\pi} \int_0^\infty \frac{\sd \om}{\om}
	\frac{\sigma(\om)}{\om^2+\om_n^2}.
    \label{eq:HE,Q^2=0}
    \eeq
    A question arises as to how to interpret the right-hand side of this equation for $\omega_n=0$.
The rule that applies to all cases of interest in the quark-gluon plasma context
    is that one should simply interpret it as being zero for $\omega_n=0$.
    In an interacting theory\footnote{Eq.\ (\ref{eq:HE,Q^2=0}) was tested~\cite{Meyer:2018xpt} in the strongly interacting super-Yang-Mills theory
    using the spectral function derived via the AdS/CFT correspondence~\cite{CaronHuot:2006te}.},
    where we expect $\sigma(\omega)/\omega$ to be finite at $\omega\to0$,
the right-hand side also vanishes if one takes its limit for $\omega_n\to0$;
however, in the theory of free thermal quarks where $\sigma(\omega)/\omega\propto  \delta(\omega)$,
the right-hand side would not vanish in that limit. This issue amounts to the same question as to whether
the function $H_E(\om_n)$, analytically continued to all frequencies, is continuous at argument zero (see~\cite{Meyer:2011gj}, sec.\ 2.3).

The integration kernels multiplying the function $\sigma(\om)/\omega$ 
in Eqs.~(\ref{eq:HEn-HEr,disprel}) and (\ref{eq:HE,Q^2=0})
are shown in Fig.~\ref{fig:kernel,cont}.
One can notice the different characteristics of these curves,
the integrand of Eq.~(\ref{eq:HE,Q^2=0}) being sensitive to soft photon emission,
while the kernel for $H_E(\omega_2)-H_E(\omega_1)$ tends to zero for $\omega/T\to 0$.
This difference is significant, since the slope of $\sigma(\om)$ at the origin gives 
access to the charge diffusion coefficient $D$~\cite{Ghiglieri:2016tvj,Ce:2020tmx,Ce:2022fot},
\beq\label{eq:Dfromsigma}
\lim_{\om\to0} \frac{\sigma(\om)}{\om} = 2 D \chi_s,
\eeq
where
\beq
\chi_s \equiv  \int d^4x\; \langle J_0(x) J_0(0)\rangle
\eeq
is the static charge susceptibility. We recall that for free massless quarks, $\chi_s = (\sum_f Q_f^2)\frac{N_c}{3}T^2$.

\begin{figure}[t]
\begin{center}
\includegraphics[scale=0.80]{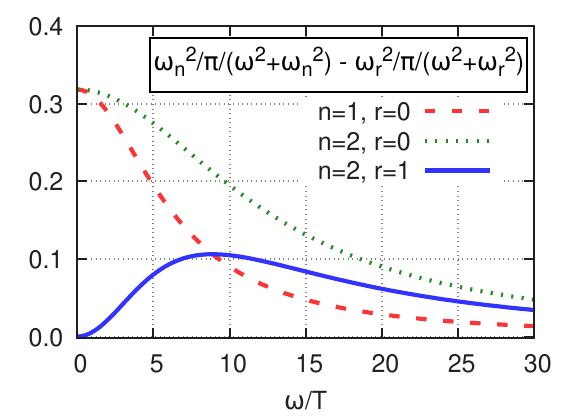}
\includegraphics[scale=0.63]{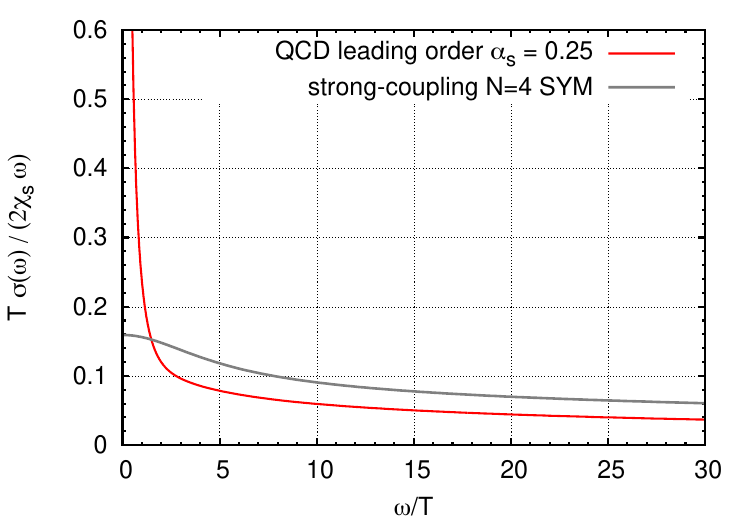}
\caption{
{\bf Left}: The integration kernels of Eqs.~(\ref{eq:HEn-HEr,disprel}) and (\ref{eq:HE,Q^2=0})
multiplying the function $\sigma(\om)/\om$.
{\bf Right}: The spectral function normalized by $2\chi_s \om/T$ at lightlike kinematics 
in QCD at complete leading order according to Ref.~\cite{Arnold:2001ba,Arnold:2001ms}, and in the strongly-coupled 
$\mathcal{N}=4$ SYM theory~\cite{Caron-Huot:2006pee}.
\label{fig:kernel,cont}
}
\end{center}
\end{figure}

\subsection{Lattice subtractions}
\label{sec:subtractions}
The lattice regulator breaks Lorentz symmetry: short-distance contributions to
$H_E(\omega_n)$ emerge which would hamper the continuum extrapolation.
It is therefore necessary to develop suitable lattice representations of
$H_E(\om_n)$.
In fact, for discretizations of the correlator which use two local currents or
two exactly-conserved lattice currents, such subtractions are required to
cancel divergences arising from contact terms and have a well-defined
continuum limit, see Sec.~\ref{sec:lat.defs.} for further discussion.
This can be achieved in several ways, for instance by subtracting the vacuum
lattice correlator obtained at the same bare parameters, as was proposed
in Ref.~\cite{Meyer:2018xpt} or by subtracting a thermal correlator having
the same short-distance properties~\cite{Meyer:2021jjr}. 

The basic observation is that the Fourier-transform of a static screening correlator 
at lightlike momentum vanishes in the continuum.
Indeed, the restriction of the polarisation tensor to spatial components 
in the static sector has the form
\beq
\Pi_{ij}(\bold{p}) \equiv  \int_0^\beta \sd x_0 \int \sd^3x\; e^{i\bold{p} \bold{x}} \; \langle J_i(x) \,J_j(0)\rangle
=  (p_i p_j - \delta_{ij} \bold{p}^{2}) \Pi(\bold{p}^{2})
	\eeq
familiar from the vacuum polarisation. From here and from the absence of a pole in $\Pi$ at $\bold{p}^{2}=0$
follows immediately the property
\beq
\int_{-\infty}^{\,\infty} \sd x_3\; {G^{\sT}_{\rm{st}}(p,x_3)} \;\eexp^{-p x_3} = 0, \quad \forall p\in\mathbb{R}.
\eeq
Thus, generalizing the estimator proposed in~\cite{Meyer:2021jjr},
we consider a subtraction involving the static screening correlator
at momentum $p$,
\begin{align}
	H_{E,\,p}(\om_n) &= -\int_0^\beta \sd x_0 \int \sd^3 x\,
	\left( {\eexp^{\om_n\, (\I x_0 - x_3)}} - {\eexp^{p\, (\I x_2 - x_3)}} \right)\,
	\langle J_1(x) J_1(0) \rangle \nonumber \\
	&= \int_{-\infty}^{\,\infty} \sd x_3\, 
	\Big[ {G^{\sT}_{\rm{ns}}(\om_n,x_3)} \eexp^{-\om_n x_3} - {G^{\sT}_{\rm{st}}(p,x_3)} \eexp^{-p x_3} \Big]\,.
	\label{eq:HEsub,cont,def}
\end{align}
We note again that in the case of the static (st) transverse channel screening 
correlator (defined in Eq.~(\ref{eq:GT,st,cont})), the momentum $p$ is inserted
into a spatial direction (here $x_2$) orthogonal to $x_3$ and to the directions
corresponding to the Lorentz indices of the currents (here $x_1$).
The case of $p=\omega_n$~\cite{Meyer:2021jjr} has the special property that $H_{E,\,\omega_n}(\om_n)$
vanishes identically at $T=0$, correctly reproducing the continuum property that
$H_E(\omega_n)$ vanishes in the vacuum. This property is expected to reduce discretisation errors
of $H_{E,\,\omega_n}(\om_n)$ at non-zero temperature, an expectation that is confirmed in lattice perturbation theory
(see App.~\ref{app:lattpt}).

We remark that even more general subtractions
\beq
	H_{E,\,{\bf p},\,{\boldsymbol\alpha}}(\om_n) = \int_{-\infty}^{\infty} \sd x_3
	\Big[ G_{\rm{ns}}^\sT(\om_n, x_3) \eexp^{-\om_n x_3} 
		  - \sum_{i} \alpha_i G_{\rm{st}}^\sT(p_i, x_3) \eexp^{-p_i x_3} \Big]
	\label{eq:HEsub,cont,def,gen}
\eeq
are possible, i.e.\ one can subtract a general linear combination of static 
screening correlators at different values of $p_i$.
Eq.~(\ref{eq:HEsub,cont,def}) is a special case of Eq.~(\ref{eq:HEsub,cont,def,gen})
with $\alpha_1=1$, $p_1=p$ and $\alpha_{i>1}=0$.
When evaluated at different values of $p$, the results for $H_{E,\,p}(\om_n)$
in Eq.~(\ref{eq:HEsub,cont,def}) do not agree with each other in general 
on a finite lattice, but they have to match after taking the continuum limit.
The same holds for the quantity in Eq.~(\ref{eq:HEsub,cont,def,gen}),
i.e.\ choosing different $\alpha_i$ coefficients and subtracting the static screening
correlators evaluated at different momenta, the results for 
$H_{E,\,{\bf p},\,\boldsymbol\alpha}(\om_n)$ differ at a finite 
lattice spacing, but have to agree in the continuum.
One can exploit these observations
and propose subtractions that may have more tractable integrands and/or reduced
cutoff effects toward the continuum.

Similarly, a direct probe of the difference of $H_E(\om_n)$ and $H_E(\om_r)$
on the lattice is provided by the discretized version of
\begin{align}
	[H_E(\om_n)- \varepsilon H_E(\om_r)]_{{\bf p},\,\boldsymbol\alpha} =
	\int_{-\infty}^{\infty} \sd x_3
    \Big[ G_{\rm{ns}}^\sT(\om_n, x_3) \eexp^{-\om_n x_3}
		 &-\varepsilon\, G_{\rm{ns}}^\sT(\om_r, x_3) \eexp^{-\om_r x_3} \nonumber \\
		 &- \sum_{i} \alpha_i G_{\rm{st}}^\sT(p_i, x_3) \eexp^{-p_i x_3} \Big].
    \label{eq:HEn-HEr,cont,def,gen}
\end{align}
Again, the contribution coming from the static screening correlators in Eq.~(\ref{eq:HEn-HEr,cont,def,gen})
vanishes in the continuum.

We will explicitly investigate the more general subtracions for the extraction 
of $H_E(\omega_2)$, i.e. in the second Matsubara sector, as well as for the difference 
$H_E(\omega_2)-H_E(\omega_1)$ in Sec.~\ref{sec:res,HE2}. 
Additionally, alternative subtractions concerning $H_E(\om_1)$ and $H_E(\om_2)$ will be exploited in Apps.~\ref{app:alter_sub,HE1}~and~\ref{app:alter_sub,HE2}, respectively.
In the rest of the paper we use the notation $H_E$ omitting the extra subscripts 
to indicate the standard subtraction with $\alpha_1=1$, $\alpha_{i>1}=0$ and $p=p_1=\omega_n$.

\subsection{Probing the virtuality dependence of $H_E$}
\label{sec:dHEdQ2}

Until this point, the above discussion focused on observables at vanishing virtuality,
since this is the relevant kinematics for the photon emissivity.
We note that $H_E$ can also be evaluated for a given non-vanishing virtuality, which
can be helpful to understand the behavior of the low-mass dilepton rate~\cite{Ce:2020wgg}.
In order to investigate the effect of introducing a small non-vanishing $Q^2$, 
we can calculate the derivative of $H_E$ with respect to $Q^2$, evaluated at $Q^2=0$.
For that, we start from the definition given in Ref.~\cite{Ce:2020wgg}
\begin{align}
	H_E(\om_n, Q^2) &= - \int_0^\beta \sd x_0\, \int \sd^3 x\,\, \eexp^{\I \om_n x_0}\,\, \eexp^{\sqrt{\om_n^2 - Q^2}x_3}\,\, \langle J_1(x) J_1(0) \rangle \nonumber \\
	&= \int_{-\infty}^{\infty} \sd x_3\, G^\sT_{E, {\rm ns}}(\om_n, x_3)\,\, \eexp^{\sqrt{\om_n^2-Q^2} x_3}.
	\label{eq:HE@Q^2,cont}
\end{align}
By expanding around $Q^2=0$ for $\om_n>0$ we obtain
\begin{align}
	H_E(\om_n,Q^2) &= H_E(\om_n,0) + Q^2\, \Bigg[\frac{\sd H_E(\om_n,Q^2)}{\sd Q^2}\Bigg]_{Q^2=0} + \mathcal{O}(Q^4) \nonumber \\
	&= H_E(\om_n,0) - \frac{Q^2}{2 \om_n} \int_{-\infty}^{\infty} \sd x_3\,\, x_3\,\, G^\sT_{E, {\rm ns}}(\om_n, x_3)\,\, \eexp^{\om_n x_3} + \mathcal{O}(Q^4).
\end{align}
Evaluating the derivative in the free theory, we can use the asymptotic formula
$G^\sT_{E, {\rm ns}}(\om_n, x_3) \overset{x_3 \to \infty}{\sim}\, \eexp^{-\om_n |x_3|}/x_3^2$ (see Ref.~\cite{Brandt:2014uda}, Eq.\ (3.11)).
With this we find that in the free theory, the $\mathcal{O}(Q^2)$ term contains an infrared-divergent term,
\beq\label{eq:logQ2wc}
	H_E(\om_n,Q^2) = H_E(\om_n,0) + c_n\, Q^2 \log(Q^2/\om_n^2) + \mathcal{O}({Q^2})
\eeq
with $c_1 = N_c/(32\pi^2)$. This coefficient can be obtained by introducing a cutoff $1/Q$ for the $x_3$-integral.

In order to ensure that we have a definition that is ultraviolet-finite,
we remove the divergence present at finite $Q^2$ by subtracting $ H_E(0, Q^2)$,
which does not change the value at $Q^2=0$ of the function we are Taylor-expanding because $H_E(0,0)=0$.
Thus we evaluate
\begin{align}
	\frac{\sd}{\sd Q^2} \left[ H_E(\om_n, Q^2) - H_E(0, Q^2) \right]_{Q^2=0} &= -\int_{0}^{\infty} \sd x_3\, \left[ \frac{x_3}{\om_n} \sinh(\om_n x_3) G^\sT_{{\rm ns}}(\om_n, x_3) - x_3^2 G^{\rm T}_{\rm st}(0, x_3) \right].
	\label{eq:dHEdQ2,cont}
\end{align}
In Eq.~(\ref{eq:dHEdQ2,cont}), $G_{\rm st}^\sT(0,x_3)$, which denotes the static, zero-momentum
screening correlator, does not yield an infrared-enhanced contribution.

\subsection{Lattice observables}
\label{sec:lat.defs.}

In this Section, we introduce the lattice observables that we have investigated.
We define the bare local and the conserved vector current as
\beq
	V_\mu^\mathrm{L}(x) = \widebar \Psi(x)\frac{\tau_3}{\sqrt{2}}\gamma_\mu\Psi(x),
	\label{eq:Vloc,bare}
\eeq
and
\beq
	V_\mu^\mathrm{C}(x) = \frac{1}{2} \left[ \widebar{\Psi}(x+a\hat\mu) (1 + \gamma_\mu) U^\dag_\mu(x) \frac{\tau_3}{\sqrt{2}}\Psi(x) - \widebar{\Psi}(x)(1 - \gamma_\mu) U_\mu(x)\frac{\tau_3}{\sqrt{2}} \Psi(x+a\hat\mu) \right],
	\label{eq:Vcons,bare}
\eeq
respectively, where $\Psi=(u,d)^\top$ represents the isospin doublet of 
mass-degenerate quark fields and $\tau_3$ is the diagonal Pauli matrix.
In other words, we focus on the isovector flavour combination, the current being normalized according to $\sum_f Q_f^2 = 1$.
It is in that normalization that our results for $H_E(\om_n)$ are given. For an estimate of the physical photon emissivity assuming
SU(3) flavour symmetry among the $(u,d,s)$ quarks, one must include the factor $\sum_f Q_f^2 = (2/3)^2+(-1/3)^2+(-1/3)^2 = 2/3$.

Using Eqs.~(\ref{eq:Vloc,bare}) and~(\ref{eq:Vcons,bare}), we define the following bare, non-static
screening correlators:
\beq
	G^{\alpha \beta}_{{\rm ns},\,\mu\nu}(\om_n, x_i) = -a^3 \hspace{-0.4cm}\sum_{\substack{x_j\\ j\in\{0,1,2,3\},\, j \ne i}} \hspace{-0.4cm} \eexp^{\I \om_n x_0}\, \langle V_\mu^\alpha(x) V_\nu^\beta(0) \rangle, \qquad i\in\{1,2,3\}, \qquad \alpha,\beta \in \{L, C\}
	\label{eq:Gns,lat}
\eeq
The Greek letters $\alpha$ and $\beta$ stand for the discretization of the current
at sink and source, respectively.
We call the above correlators non-static, because the momentum is inserted into
the Euclidean time-direction.
We denote these with the subscript ${\rm ns}$.
By contrast, when injecting the momentum into a spatial direction, 
$k$, -- perpendicular to direction of $x_i$ --, we got the bare,
static screening correlators at finite momentum,
\beq
	G^{\alpha \beta}_{{\rm st},\,\mu\nu}(\om_n, x_i) = -a^3 \hspace{-0.4cm}\sum_{\substack{x_j\\ j\in\{0,1,2,3\},\, j \ne i}} \hspace{-0.4cm} \eexp^{\I \om_n x_k}\, \langle V_\mu^\alpha(x) V_\nu^\beta(0) \rangle, \quad i,k\in\{1,2,3\},\,\quad i\ne k, \quad \alpha,\beta \in \{L, C\}.
	\label{eq:Gst,lat}
\eeq
After choosing a particular spatial decay direction (direction of the correlator separation),
$i$, we obtain the screening correlators in the transverse channel by choosing
$\mu=\nu$ orthogonal to this decay direction.
We note here that we do not discriminate the notation used for the continuum or
the lattice observables, i.e.\ we use capital $G$ for the lattice and for the
continuum screening correlators as well.

We also specify the correlator at a momentum transverse to both $\mu=\nu$ and $i$.
Therefore, we have in total six possible combinations of the decay direction and
of the Lorentz indices of the currents for the non-static screening correlator
and also six combinations of the decay direction, the Lorentz indices of the currents
and the momentum inserted for the static screening correlator.
We average over these different screening correlators measured on the same
configuration.
Moreover, the local-conserved and conserved-local discretizations can be transformed
into each other, using Cartesian coordinate reflections. 
Therefore, we average these two and refer to this averaged correlator with the
superscript LC in the following.
We renormalize the correlators by multiplying by $Z_V(g_0^2)$ whenever the local
vector current is included using the vector current renormalization constant
from Ref.~\cite{DallaBrida:2018tpn}.
Instead of the electromagnetic current, we use the isovector vector current
whereby disconnected contributions are absent.

As mentioned already at the beginning of Sec.~\ref{sec:subtractions}, 
one has to pay special attention when formulating a lattice estimator for $H_E$, 
since a naive implementation of Eq.~(\ref{eq:HE,intGscr}) on the lattice, 
could lead to ultraviolet divergences.
A crucial point at small separation $x$ is the removal of the $x$-independent part
multiplying the expectation value of the product of currents in Eq.~(\ref{eq:HE,intGscr}).
The simplest way of doing this is to subtract the static screening correlator
at vanishing momentum, or to subtract it evaluated at the same finite momentum $\om_n$,
\begin{align}
	H_E^{\rm (sub)}(\om_n)/T^2 = \frac{a}{L_t} \sum_{x_3=a}^{L_s/2-a} h(\om_n, x_3) + \frac{a}{2 L_t}\Big( h(\om_n, 0) + h(\om_n, L_s/2) \Big)
	\label{eq:HEsub,lat,def}
\end{align}
where we used the trapezoid formula when discretizing Eq.~(\ref{eq:HEsub,cont,def})
with $p=\om_n$ and $L_t$ ($L_s$) stands for the temporal (spatial) size of the lattice.
The lattice spacing is deenoted by $a$ and $h(\om_n, x_3)$ is the integrand,
\beq
	h(\om_n, x_3) = \frac{2}{T^3} \Big( G_{\rm ns}^\sT(\om_n,x_3) - G_{\rm st}^\sT(\om_n,x_3) \Big) \cosh(\om_n x_3).
	\label{eq:HEsub,lat,intnd0}
\eeq

We note that $G^\sT_{\rm ns}$ as well as $G^\sT_{\rm st}$ are negative in our conventions
and since in absolute value we found $G^\sT_{\rm ns}$ larger than $G^\sT_{\rm st}$, 
$H_E^{\rm (sub)}(\om_n)/T^2$ will also be negative.
From now on we leave the (sub) superscript from $H_E^{\rm (sub)}(\om_n)$, 
and similarly to the correlators, we do not discriminate between the continuum 
and lattice observables, i.e.\ we use the same symbol.

Analogously as in the case of $H_E(\om_n)$, one can obtain the lattice formula 
for the derivative of $H_E(\om_n, Q^2)$ w.r.t. $Q^2$ by applying the trapezoid formula 
to Eq.~(\ref{eq:dHEdQ2,cont}), but where $h(\om_n, x_3)$ of the right-hand side of
Eq.~(\ref{eq:HEsub,lat,def}) is replaced by
\beq
	h_{Q^2}(\om_n, x_3) = - \frac{1}{T^3} \Big( \,\frac{x_3 T}{\om_n/T}\, \sinh(\om_n x_3)\, G^\sT_{\rm{ns}}(\om_n, x_3)
		- (x_3 T)^2 G^\sT_{\rm{st}}(0, x_3) \Big).
	\label{eq:dHEdQ2,intnd}
\eeq
We note that this subtraction involves the completely static --- zero-momentum ---
screening correlator, $G_{\rm{st}}^\sT(p=0, x_3)$.

\subsection{Simulation details}
\label{sec:simul}
To calculate the screening correlators which enter the expression~(\ref{eq:HEsub,lat,def})
for $H_E$, we used three ensembles generated at the same temperature $T \sim 250$ MeV in the
high-temperature phase.
We employ two-flavor O($a$)-improved dynamical Wilson fermions and the plaquette gauge action;
further details regarding the lattice action we used can be found in Ref.~\cite{Fritzsch:2012wq}.
The configurations for the W7 ensemble have been generated using the openQCD-1.6
package and the ones of O7 and most of those of X7 using openQCD-2.0~\cite{Luscher:2012av}.
512 configurations of the X7 ensemble were generated using the MP-HMC algorithm~\cite{Hasenbusch:2001ne} in the implementation described in Ref.~\cite{Marinkovic:2010eg}.
The pion mass in the vacuum is around $m_\pi \approx 270$ MeV~\cite{Fritzsch:2012wq,Engel:2014cka}, 
and the lattice spacings are in the range of 0.033--0.05 fm~\cite{Fritzsch:2012wq,Ce:2021xgd}.
The boundary conditions are periodic in space, while those in the time direction are periodic
for the gauge field and antiperiodic for the quark fields, as required by the Matsubara formalism.

\begin{table}[t]
\begin{tabular}{cccccc}
\hline\hline
label & $6/g_0^2$ & $\kappa$ & $L_t/a$ & $N_{\rm conf}$ & $\frac{\rm MDUs}{\rm conf\phantom{_A}}$ \\
\hline
O7 & 5.5 & 0.13671        & 16          & 1500  & 20 \\ 
W7 & 5.685727\,\, & 0.136684  & 20          & 1600 &  8 \\ 
X7 & 5.827160\,\, & 0.136544  & 24          & 2012  & 10 \\ 
\hline\hline
\end{tabular}
\caption{\label{tab:ensembles} Overview of the $N_f=2$ ensembles used in this study.
Simulations were carried out at a fixed temperature of $T \approx 254\,$MeV
and fixed aspect ratio $L_s/L_t=4$, where $L_s$ ($L_t$) is the spatial (temporal)
linear size of the lattice.
The parameters given are the bare gauge coupling $g_0$, the Wilson hopping 
parameter $\kappa$, the temporal size in units of the lattice spacing $a$,
the number of configurations used $N_{\rm conf}$ and the number of molecular-dynamics
time units (MDUs) separating these configurations.
The number of point sources per configuration is 64\ in all cases.
}
\end{table}
\section{Results}
\label{sec:results}
\subsection{The integrand for obtaining $H_E$}
\label{sec:HE,intnd}

As we have shown in Sec.~\ref{sec:lat.defs.}, the crucial ingredients for calculating
$H_E$ through Eq.~(\ref{eq:HEsub,lat,def}) are the non-static and static transverse screening
correlators at finite spatial momentum.
These have been first investigated in detail in weak-coupling theory complemented
with lattice QCD simulations on a single ensemble in Ref.~\cite{Brandt:2014uda}.
In that work, however, the static screening correlators have been studied only
at vanishing momentum.

\begin{figure}[t]
\begin{center}
\includegraphics[scale=0.80]{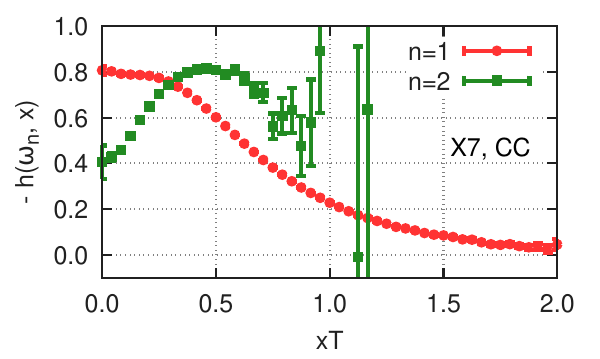}
\caption{The integrand (Eq.~(\ref{eq:HEsub,lat,intnd0})) for the calculation of $H_E(\om_n)$ on our
finest ensemble X7, using the conserved-conserved discretization of the
current-current correlator.
\label{fig:HEintnd}
}
\end{center}
\end{figure}

For the determination of $H_E$, we first analyzed the screening correlators
measured on the three ensembles.
We discarded a few outliers from the dataset (see App.~\ref{app:outliers}) and estimated
the statistical errors using jackknife resampling with 200 jackknife samples.
Then we formed the integrand (see Eq.~(\ref{eq:HEsub,lat,intnd0})), which is shown 
in Fig.~\ref{fig:HEintnd} for our finest ensemble.

As Fig.~\ref{fig:HEintnd} shows, the quantity $H_E(\om_1)$ receives dominant contributions
from the $xT \lesssim 1-1.5$ region and with the present statistics, we have good control
over the signal.
The relative error on the integrand for $H_E(\om_1)$ is below 1\% for $xT \lesssim 1.1$
on the finest ensemble shown in Fig.~\ref{fig:HEintnd}.
However, trying to evaluate $H_E$ in the $n=2$ Matsubara sector reveals that one faces
a severe, exponential signal-to-noise problem.
The variances on the non-static correlator values are an order of magnitude larger in the $n=2$
sector than for $n=1$, and multiplying by $\cosh(2 \pi T n)$ makes the situation even worse.
Therefore, we focus first on evaluating $H_E$ in the $n=1$ Matsubara sector and return
to $n=2$ only in Secs.~\ref{sec:alter_sub,HE2} and ~\ref{sec:HE2-HE1}.

\subsection{Modelling the tail of the screening correlators}
\label{sec:tail}

While the dominant contribution to $H_E(\om_1)$ comes from short distances,
the long-distance contribution is also non-negligible.
However, it is noisier, because the screening correlators are less precise at large
distances and the difference is less smooth than for short distances.
Directly performing the sum of Eq.~(\ref{eq:HEsub,lat,def}) for $H_E(\om_1)$
results in having relative errors of about $1.37, 1.50, 1.90\%$ for the LL, LC and CC
discretizations, respectively, on our coarsest and also `noisiest' ensemble.
We aim at a more precise determination of $H_E(\om_1)$ and as we will see in this
section, by proper handling of the tail these errors could be reduced 
to $0.91, 0.77, 1.03\%$, respectively.

Moreover, we have an exponentially growing weight function multiplying 
the difference of the correlators and this results in a small enhancement of the integrand
in the region $x_3 \sim L/2$.
This effect is the consequence of calculating the integrand in a finite volume.
In order to correct for it, we applied a simple model based on fits on the
screening correlators to describe the tail of the integrand.
This way we have better control over the long-distance contribution and also
we could correct for finite volume effects.
We use the fact, that the non-static screening correlators have a representation
in terms of energies and amplitudes of screening states in the following form~\cite{Meyer:2018xpt}:
\beq
	G_{\rm{ns}}(\om_r,x) \overset{x \ne 0}{=}
	\sum_{n=0}^{\infty} |A_{\rm{ns},n}^{(r)}|^2 \eexp^{-E_{\rm{ns},n}^{(r)} |x|}.
	\label{eq:ns,scrcorr,asympt}
\eeq
A similar expression holds for the static correlator:
\beq
	G_{\rm{st}}(\om_r,x) \overset{x \ne 0}{=}
	\sum_{n=0}^{\infty} |A_{\rm{st},n}^{(r)}|^2 \eexp^{-E_{\rm{st},n}^{(r)} |x|}.
	\label{eq:st,scrcorr,asympt}
\eeq
The low-lying screening spectrum can be studied using weak-coupling methods
as well~\cite{Brandt:2014uda}.
The lowest energy of a screening state in a given Matsubara sector with frequency
$\om_r$ is often called the screening mass and is denoted by $E^{(r)}_0$.
In this section, we consider only the first Matsubara sector, $\om_1 = 2 \pi T$,
therefore in the following we do not write out explicitly the momentum dependence.

Using the above formulae, our procedure to get a better handle over the
integrand is the following:
\begin{enumerate}
	\item We split the integrand, $h(x) \equiv h(\om_1, x)$ (c.f. Eq.~(\ref{eq:HEsub,lat,intnd0}))
		  into two parts
		  \beq
				h(x) = h(x)(1-\Theta(x,x_w,\Delta)) + h(x)\Theta(x,x_w,\Delta)
				\label{eq:h,split}
		  \eeq
		  using a smooth step function,
		  \beq
		  		\Theta(x,x_w,\Delta) = (1+\tanh[(x-x_w)/\Delta])/2.
				\label{eq:step_fun}
		  \eeq
		  We call the first term in Eq.~(\ref{eq:h,split}) the short-distance and the second term the long-distance
		  part of the integrand.
	\item We integrate the short-distance contribution using the trapezoidal formula.
	\item We perform single-state fits on the tails of the screening correlators using the
		  representations given in Eqs.~(\ref{eq:ns,scrcorr,asympt}) and (\ref{eq:st,scrcorr,asympt})
		  translated to a form corresponding to a periodic lattice, namely
		  \beq
				G^{\rm{(ns)}}_{\rm{ansatz}}(x) = |A_{\rm{ns},0}|^2 \cosh\big[E_{\rm{ns},0} (x-L_s/2)\big]
				\label{eq:G_ans,ns}
		  \eeq
		  and
		  \beq
				G^{\rm{(st)}}_{\rm{ansatz}}(x) = |A_{\rm{st},0}|^2 \cosh\big[E_{\rm{st},0} (x-L_s/2)\big],
				\label{eq:G_ans,st}
		  \eeq
		  for the non-static and for the static screening correlators, respectively.
		  In Eqs.~(\ref{eq:G_ans,ns}) and (\ref{eq:G_ans,st}), $L_s$ is the spatial length
		  of the lattice.
	\item Using the fit results $A_{\rm{ns},0}$ and $E_{\rm{ns},0}$ 
		  as well as $A_{\rm{st},0}$ and $E_{\rm{st},0}$, we replace 
		  the long-distance part of Eq.~(\ref{eq:h,split}) by the corresponding
		  infinite volume formula:
		  \begin{align}
			    h_{\rm{ansatz,iv}}(x)\Theta(x,x_w,\Delta) &= 
				\Big[
				|A_{\rm{ns},0}|^2 \frac{\eexp^{E_{\rm{ns},0} L_s/2}}{2} 
				\Big( \eexp^{-(E_{\rm{ns},0}+\om_1)x} + \eexp^{-(E_{\rm{ns},0}-\om_1)x}\Big) \nonumber \\
				&-
				|A_{\rm{st},0}|^2 \frac{\eexp^{E_{\rm{st},0} L_s/2}}{2} 
				\Big( \eexp^{-(E_{\rm{st},0}+\om_1)x} + \eexp^{-(E_{\rm{st},0}-\om_1)x}\Big)
				\Big]
				\Theta(x,x_w,\Delta).
				\label{eq:h_ans}
		  \end{align}
		  This we can integrate analytically using:
		  \beq
				\int \sd x\, \eexp^{-\alpha_1 x} \tanh(x-\alpha_2) = \frac{1}{\alpha_1} \eexp^{-\alpha_1 x} \Big( 2 {}_2 F_1(1, -\alpha_1/2; 1-\alpha_1/2; -\eexp^{2x-2\alpha_2}) - 1 \Big) + \rm{const},
				\label{eq:int,exptanh}
		  \eeq
		  where ${}_2F_1$ denotes the hypergeometric function.
\end{enumerate}

While the single-state fits describe the actual data well, i.e.\ with good $\chi^2$
and p-values, we note that the identification of the plateau region was not clear
in some cases, although we performed a thorough scan using all possible fit ranges
having different starting points and different lengths with 6$a$--11$a$.
Therefore, we also made an attempt to fit the data with two-state fits with or without
using priors from weak-coupling theory (c.f. Sec.~\ref{app:weak-cth}), but these
fit results were not satisfactory.
Typically on the coarsest or the coarser two ensembles, they either failed to
describe the data, gave too large errors or returned a near-zero or negative
$|A|^2$ coefficient --- which we did not constrain --- for the excited state.
When the two-state fits were able to describe the data well, the ground-state
static screening energy they returned was too small --- as we could deduce it
using the zero-momentum correlators.
Therefore we decided to stick to single-state fits.

\begin{figure}[t]
\begin{center}
\includegraphics[scale=0.84]{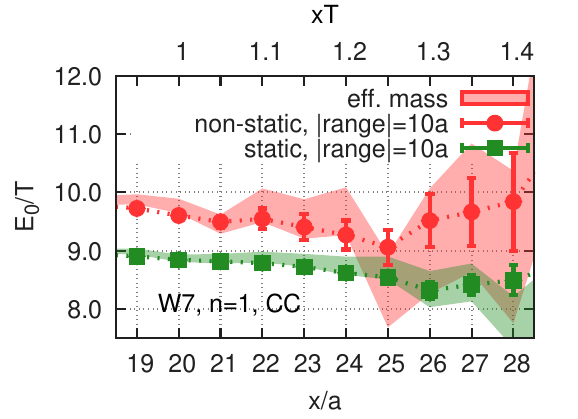}
\includegraphics[scale=0.76]{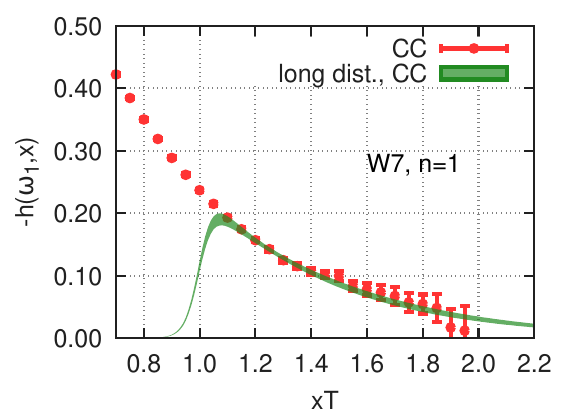}
\caption{
{\bf Left}: fitted masses, with circles and squares corresponding to $G_{\rm ns}^\sT(\om_1,x_3)$ and 
$G_{\rm st}^\sT(\om_1,x_3)$ respectively, using a fit range of ten lattice spacings.
The bands represent the 'cosh' effective masses.
{\bf Right}: the tail of the integrand needed for the calculation of $H_E$ 
at the first Matsubara frequency.
The red points are the actual datapoints on our second coarsest lattice and 
the band shows the result of the modelling.
}
\label{fig:extrmass&intnd}
\end{center}
\end{figure}

Besides fitting, we also determined the effective mass using two consecutive
correlator datapoints, by solving the algebraic equation
\beq
	\frac{G(x+a)}{G(x)}
	= \frac{\cosh\big[m_{\rm{eff}} (x+a-L_s/2)\big]}{\cosh\big[m_{\rm{eff}} (x-L_s/2)\big]}
\eeq
for $m_{\rm{eff}}$.
Here, $G(x)$ denotes the actual lattice data for the non-static or the static
screening correlators.
We found that the effective masses are in quite good agreement with the fitted masses,
but also do not show a clear and long plateau as $x$ increases,
see Fig.~\ref{fig:extrmass&intnd}, left panel.
Therefore, instead of fitting a constant, we decided to choose three representatives
from a histogram built by assigning Akaike-weights~\cite{Akaike:1973abc,Borsanyi:2020mff}
to all the fitted masses that we obtained in the most plateau-like region.
In all cases, we chose a wide region, before the noise gets too large on the fitted masses.
For instance for the W7 ensemble, we choose fit results on conserved-conserved correlator
data in the range from $x/a=21$ to 25 for both correlators
(left panel of Fig.~\ref{fig:extrmass&intnd}).
We propagate the median as well as the values near the 16th and 84th percentiles
of the histograms to the later steps of the analysis of $H_E(\om_1)$.
\subsection{Continuum extrapolation of $H_E(\om_1)$}
\label{sec:HE,cont.lim.}

\begin{figure}[t]
\begin{center}
\includegraphics[scale=0.80]{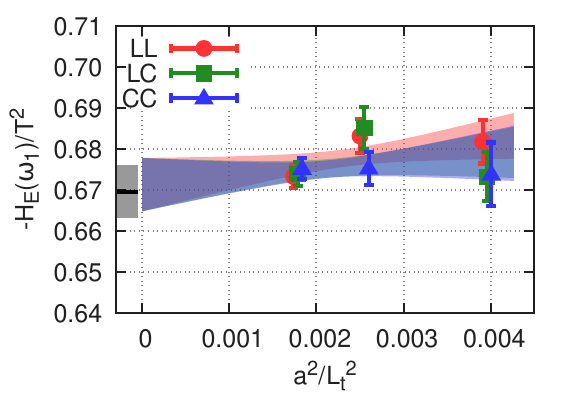}
\includegraphics[scale=0.80]{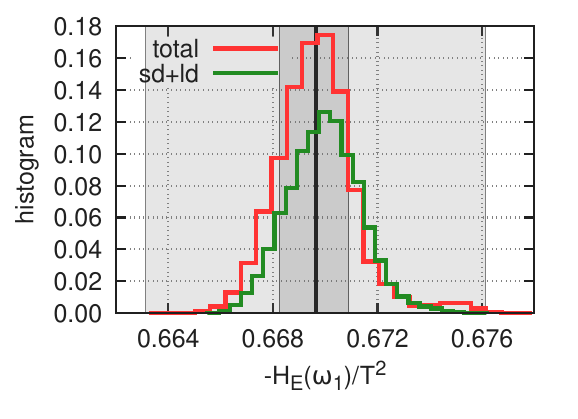}
\caption{{\bf Left}: A representative continuum limit of $-H_E(\om_1)/T^2$ obtained
  from Eqs.\ (\ref{eq:HEsub,lat,def}--\ref{eq:HEsub,lat,intnd0}).
{\bf Right}: AIC-weighted histogram of the continuum extrapolated results for $-H_E(\om_1)/T^2$ obtained with $x_w T =1.1$.
The AIC-weighted histogram of the long-distance contribution to $H_E(\omega_1)$, shifted
with the continuum result for the short-distance contribution, is also shown for 
comparison (sd+ld).
The dark-grey band shows the systematic error, and the lighter gray band represents
the total error obtained from the statistical and systematic errors added in quadrature.
}
\label{fig:HE,contlim}
\end{center}
\end{figure}

As we described in Sec.~\ref{sec:tail}, we used single-state fits
to describe the tail of the non-static and static screening correlators
and for each ensemble and discretization we built a histogram of the
fit results from the plateau region, from which we chose three representatives.
When proceeding this way for each correlators, we obtain $3\times 3=9$
possibilities for modelling the tail of the integrand, Eq.~(\ref{eq:h_ans}),
for a given ensemble and a given discretization.
We calculated $H_E$ using all these nine combinations for the tail, sorted the
results and then chose the median, the values near the 16th and near
the 84th percentile after assigning uniform weights for these slightly different
values of $H_E$.

This way we had three representative values of $H_E$ for each ensemble and discretization
that went into the next step of the analysis, which was the continuum extrapolation.
We used these in all possible combinations when performing a correlated 
simultaneous continuum extrapolation using a linear ansatz in $a^2$.
These gave $(3^3)^3=19683$ different continuum extrapolations.
We make further variations by omitting one of the coarsest datapoints 
from the extrapolation, which also lead to $3^3 \cdot 3^3 \cdot (3^2 \cdot 3)=19683$
different continuum extrapolations.
We then built an AIC-weighted histogram from using all $2 \cdot 19683$ continuum
extrapolations to estimate the systematic error.
A representative continuum extrapolation as well as the AIC-weighted histogram
are shown in Fig.~\ref{fig:HE,contlim}, left and right panel, respectively.

The transition to the modelled tail has been introduced smoothly by using
a smooth step function of Eq.~(\ref{eq:step_fun}) and we investigated the effect
of choosing different switching points, $x_w$, in the range $x_w T$=0.9--1.3.
We found that the results were stable against these choices,
see Fig.~\ref{fig:HE,contlim,comp}.

Our final result for $H_E$ in the first Matsubara sector is
\beq
	H_E(\om_1)/T^2 = -0.670(6)_{\rm{stat}}(1)_{\rm{sys}}.
	\label{eq:HE1,final,cont}
\eeq

\begin{figure}[t]
\begin{center}
\includegraphics[scale=0.80]{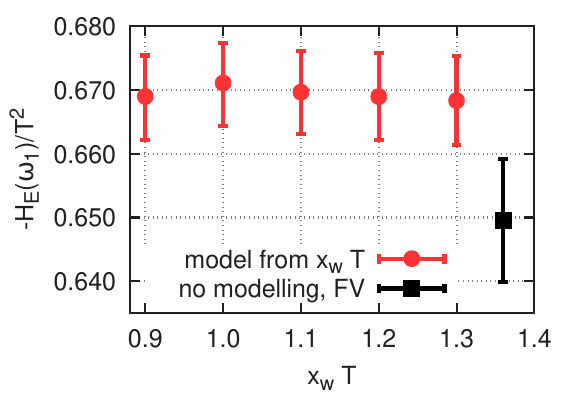}
\caption{
Comparison of continuum results for $-H_E(\om_1)/T^2$ using different starting points for the modelling.
}
\label{fig:HE,contlim,comp}
\end{center}
\end{figure}

\subsection{Continuum extrapolation of the $Q^2$-derivative of $H_E(\om_1,Q^2)$ at $Q^2=0$}
\label{sec:dHEdQ2,cont.lim.}

\begin{figure}[t]
\begin{center}
\includegraphics[scale=0.78]{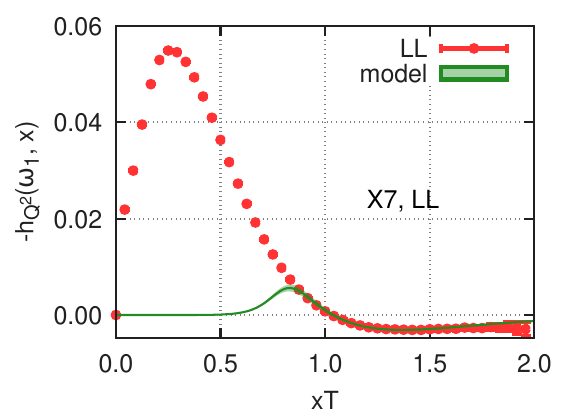}
\includegraphics[scale=0.80]{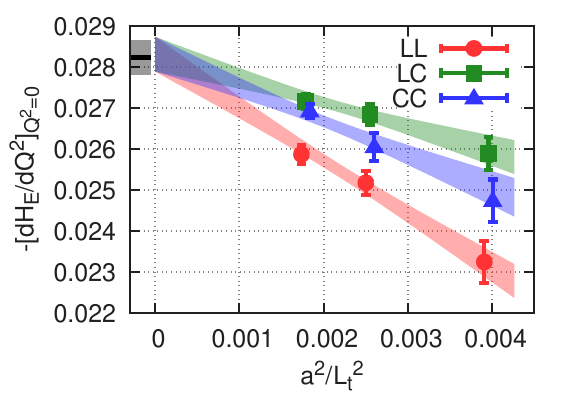}
\caption{
{\bf Left}: The integrand, $h_{Q^2}(\om_1,x)$, defined in Eq.~(\ref{eq:dHEdQ2,intnd}) for
the derivative of $H_E(\om_1,Q^2)$ with respect to $Q^2$ at $Q^2=0$.
{\bf Right}: Representative continuum extrapolation of $\sd [ H_E(\om_1,Q^2) - H_E(0,Q^2)]/\sd Q^2$ at $Q^2=0$.
}
\label{fig:dHEdQ2}
\end{center}
\end{figure}

In order to retrieve information about the $Q^2$-dependence of our
observable, we evaluate the $Q^2$-derivative of the difference, 
$H_E(\om_1,Q^2) - H_E(0, Q^2)$, as it was discussed in Sec.~{\ref{sec:dHEdQ2}}.
The continuum observable is defined in Eq.~(\ref{eq:dHEdQ2,cont}), and the
corresponding integrand on the lattice is introduced in Eq.~(\ref{eq:dHEdQ2,intnd}).
It is interesting to have a look on this integrand --- left panel of Fig.~\ref{fig:dHEdQ2} ---
which is more pronounced at short distances than the integrand $h(\om_1,x)$, that
we had shown for $H_E(\om_1)$ in Fig.~\ref{fig:HEintnd}.
It starts from zero quadratically in $x$, and after having a peak, it crosses
zero around $xT \sim 1$, but the long-distance contribution is much more suppressed
than it was for $H_E(\om_1)$.

For the continuum extrapolation we applied a similar proceduce as for $H_E(\om_1)$
discussed in Sec.~\ref{sec:HE,cont.lim.}.
Our final continuum estimate is
\beq
\frac{\sd}{\sd Q^2} \left[ H_E(\om_n, Q^2) - H_E(0, Q^2) \right]_{Q^2=0} = -0.0282(4)_{\rm{stat}}(1)_{\rm{sys}}.
\eeq
We remark that the result is on the order of $-N_c/(2\pi)^2$ and 
does not exhibit any strong infrared enhancement as would be expected at very weak coupling (see Eq.\ (\ref{eq:logQ2wc})).
        
\subsection{Continuum extrapolation of the screening masses}
\label{sec:E0,cont}

\begin{figure}[t]
\begin{center}
\includegraphics[scale=0.80]{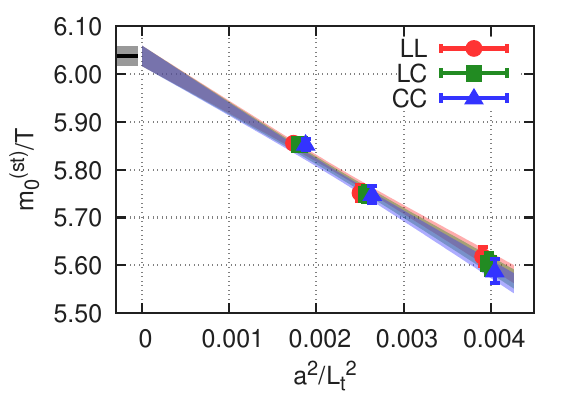}
\caption{
Continuum extrapolation of the static screening mass obtained using 
the screening correlator at vanishing momentum.
}
\label{fig:E0,n=0}
\end{center}
\end{figure}

As we have already mentioned in Sec.~\ref{sec:cont.defs.}, the screening masses
extracted from the correlators we investigate here can also be determined using
weak-coupling theory.
The first relevant study in this direction was Ref.~\cite{Brandt:2014uda}, which
also compared the leading-order (LO) and next-to-leading order (NLO) screening 
masses to lattice results.
However, Ref.~\cite{Brandt:2014uda} investigated only screening correlators at
finite temporal or at vanishing momentum.\footnote{
We note, that the terminology that we use in this paper is different from that
of Ref.~\cite{Brandt:2014uda}, the 'static ($k_n=0$)' results of that work
correspond to our zero-momentum, i.e.\ $n=0$ results.
}
Moreover, the lattice investigation has used only coarse ensembles without taking
the continuum limit.

In this work, we extend this first investigation in several aspects:
we investigate also screening masses obtained from screening correlators 
at finite spatial momentum besides the ones obtained at finite temporal
or at vanishing momentum.
We accumulated much more configurations and performed more measurements,
enabling us to improve the signal-to-noise ratio at the tails of the correlators.
Finally, we also extrapolate our results to the continuum using three ensembles,
from which the finest has a lattice spacing about 2/3 the lattice spacing
of Ref.~\cite{Brandt:2014uda}.

\begin{figure}[t]
\begin{center}
\includegraphics[scale=0.80]{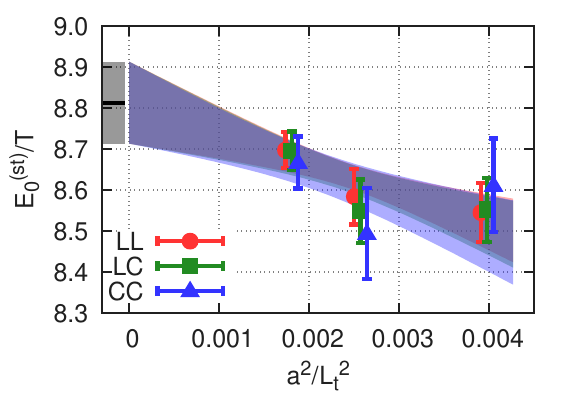}
\includegraphics[scale=0.80]{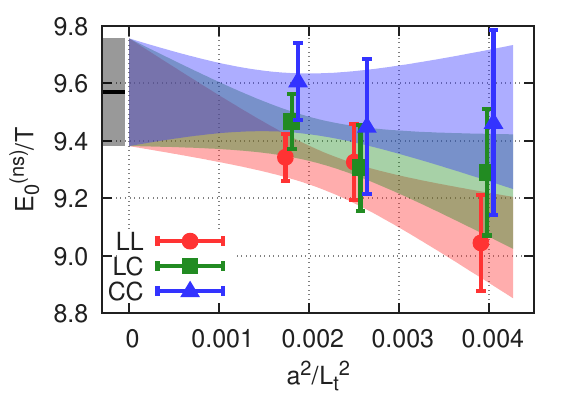}
\caption{
  Continuum extrapolation of the lowest screening mass associated with the correlators $G^\sT_{\rm st}(\omega_1,x_3)$
and $G^\sT_{\rm ns}(\omega_1,x_3)$, left and right panel, respectively.
}
\label{fig:E0,n=1,contlim}
\end{center}
\end{figure}

We start by discussing the zero and finite spatial momentum results.
We performed single-state fits --- discussed also in Sec.~\ref{sec:tail} ---
using the fit ansatz given in Eq.~(\ref{eq:G_ans,st}).
The screening correlators at vanishing momentum are more precise
above $xT\sim 1.2$ than at the first spatial Matsubara momentum,
making it possible to determine the screening masses at 0.15--0.45\% precision
depending on the ensembles and discretizations.
By using the dispersion relation, we could compare the extracted masses
at vanishing momentum to those at momentum $2 \pi T$.
With this comparison, we observe that the screening masses determinded
via the dispersion relation, $\sqrt{m_0^{(\rm{st})\,2} + (2\pi T)^2}$, give a
slightly smaller value than the corresponding result at the first
Matsubara momentum.

Since the single-state fitting procedure starting at later and later
datapoints provides fit results converging to an asymptotic value
more reliably than in the cases with finite momentum and the errors
are comparable in the plateau region, we chose only one representative
for each ensemble and discretization and performed the continuum
extrapolation linear in $a^2$ using that.
We also assigned a systematic error to the continuum extrapolation by
removing one of the discretizations on the coarsest ensemble.
The continuum extrapolation using all datapoints is shown in Fig.~\ref{fig:E0,n=0}. 
Our continuum estimate having about 0.37\% error is
\beq\label{eq:m0st}
	m_0^{(\rm st)}/T=6.04(2)_{\rm{stat}}(1)_{\rm{sys}}.
\eeq
Using this result and the dispersion relation, an estimate for 
static screening mass at (spatial) momentum $\om_1= 2 \pi T$ is
\beq
	E_{0,\rm{dr}}^{\rm{(st)}}/T = \sqrt{(m_0^{\rm (st)}/T)^2 + (2\pi)^2} = 8.72(2)_{\rm{stat}}(1)_{\rm{sys}}.
	\label{eq:E0,n0+dr->n1}
\eeq

\begin{figure}[t]
\begin{center}
\includegraphics[scale=0.80]{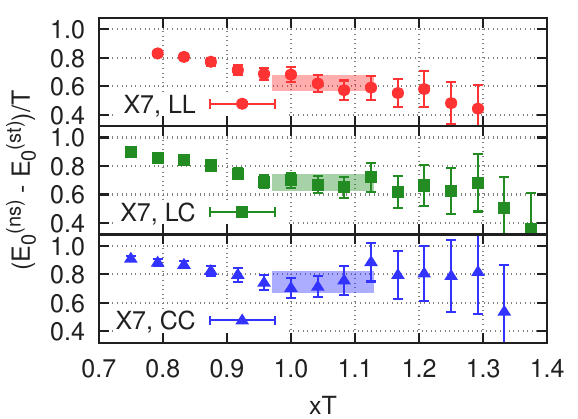}
\includegraphics[scale=0.80]{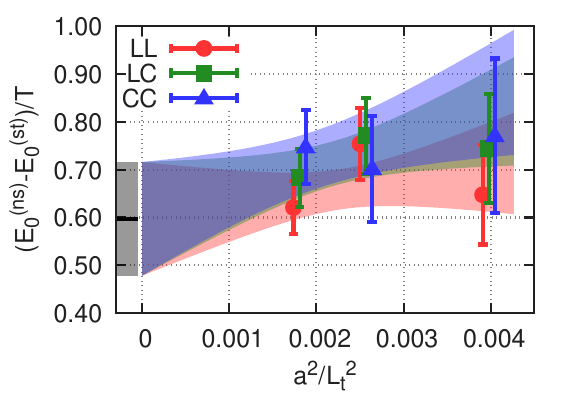}
\caption{
{\bf Left}: Fit results for the gap between the ground-state non-static and static
screening masses, respectively associated with the correlators $G_{\rm ns}^\sT(\om_1,x_3)$ $G_{\rm st}^\sT(\om_1,x_3)$,
and the window-smeared results shown by the bands.
{\bf Right}: Continuum extrapolation of the difference of the ground-state non-static and static screening masses.
}
\label{fig:Ens-Est,X7}
\end{center}
\end{figure}

The static screening masses being directly available at this momentum,
we can extra\-po\-late those to the continuum and see how
close we get to the estimate given in Eq.~(\ref{eq:E0,n0+dr->n1}).
Therefore, we calculated a weighted average of the masses
obtained by fitting the static screening correlator.
We performed the averaging using the difference of smooth step functions
--- introduced in Eq.~(\ref{eq:step_fun}) --- with parameters $x_w T=0.975$
and $x_w T=1.125$, i.e.\ the averaging window has a width about 0.15 and
was centered at 1.05.
$T \Delta$ was chosen to be 0.01.
We made similar variations as before, leaving out one datapoint from
the coarsest ensemble and arrived at the continuum estimate:
\beq
	E_0^{\rm{(st)}}/T= 8.81(10)_{\rm{stat}}(2)_{\rm{sys}}
	\label{eq:E0st,n1}
\eeq
The continuum extrapolation is shown in the left panel of Fig.~\ref{fig:E0,n=1,contlim}.
The result in Eq.~(\ref{eq:E0st,n1}) is in good agreement with the estimate
based on using the $n=0$ screening mass and the dispersion relation,
Eq.~(\ref{eq:E0,n0+dr->n1}).

Using the same procedure for the continuum extrapolation of the
non-static screening mass as in the case of the static, we obtained
the following result in the first Matsubara sector:
\beq
	E_0^{\rm{(ns)}}/T= 9.57(19)_{\rm{stat}}(11)_{\rm{sys}}.
	\label{eq:E0ns,n1}
\eeq
The continuum extrapolation is shown in the right panel of Fig.~\ref{fig:E0,n=1,contlim}.

As a crosscheck, we also investigated the possibility of determining
the non-static screening mass using the ratio of the non-static
and static screening correlators.
By making use of the following approximate formula,
\beq
    \frac{|A_n^{\rm{(ns)}}|^2}{|A_n^{\rm{(st)}}|^2}
	\frac{\cosh[E_0^{\rm{(ns)}}( x-L/2 )]}{\cosh[E_0^{\rm{(st)}}( x-L/2 )]}
    \approx C \eexp^{-(E_0^{\rm{(ns)}}-E_0^{\rm{(st)}}) x}, \quad (x\ll L/2),
	\label{eq:ratio0i}
\eeq
we fitted the ratio of the correlators using the ansatz on the
right-hand side of this equation.
After performing the averaging over the obtained differences
$(E_0^{\rm{(ns)}}-E_0^{\rm{(st)}})/T$, using the window function
with the same parameters as in the case of the static screening mass
analysis, the resulting continuum extrapolation is shown in
Fig.~\ref{fig:Ens-Est,X7}, right.
The outcome of the averaging using the window function centered at 1.05
is shown for the three discretizations on the finest ensemble
in Fig.~\ref{fig:Ens-Est,X7}, left panel.
The continuum estimate for the gap between the non-static and static
screening masses at $n=1$:
\beq
	(E_0^{\rm{(ns)}}-E_0^{\rm{(st)}})/T=0.60(12)_{\rm{stat}}(2)_{\rm{sys}}.
\eeq

Adding this to the value of the static screening mass determined directly
from fitting the data, Eq.~(\ref{eq:E0st,n1}), we obtain
$8.81(10)_{\rm{stat}}(2)_{\rm{sys}}+0.60(12)_{\rm{stat}}(2)_{\rm{sys}} = 9.41(16)_{\rm{tot}}$.
Adding it to the static screening mass estimate using the continuum dispersion
relation, Eq.~(\ref{eq:E0,n0+dr->n1}), 
we get $8.72(2)_{\rm{stat}}(1)_{\rm{sys}}+0.60(12)_{\rm{stat}}(2)_{\rm{sys}} = 9.32(12)_{\rm{tot}}$.
Both results are in good agreement with the result of Eq.~(\ref{eq:E0ns,n1}).

\section{Results concerning the $n=2$ Matsubara sector}
\label{sec:res,HE2}

In this section, we summarize the determination of $H_E(\om_2)/T^2$, i.e.\ in the second Matsubara sector, and of the difference $H_E(\om_2)-H_E(\om_1)$ using
the more general subtractions given in Eqs.~(\ref{eq:HEsub,cont,def,gen}) and (\ref{eq:HEn-HEr,cont,def,gen}).
More detailed discussion and further results applying these subtractions
can be found in Appendices~\ref{app:alter_sub,HE1}~and~\ref{app:alter_sub,HE2}.

\subsection{Continuum extrapolation of various estimators for $H_E(\om_2)$}
\label{sec:alter_sub,HE2}

As was discussed in Sec.~\ref{sec:subtractions}, when determining $H_E(\om_n)$, one is
not limited to subtract the static screening correlator with the same momentum
as the non-static screening correlator, but other choices are also possible.
We referred to the momentum of the static screening correlator by adding a subscript
$p$ and wrote $H_{E,p}(\om_n)$ for the estimator in Eq.~(\ref{eq:HEsub,cont,def}).
This subscript was not used in other sections of the paper, since we have subtracted
the static screening correlator with $p=\om_1$ when determining $H_E$ in the first
Matsubara sector in Secs.~\ref{sec:HE,intnd}--\ref{sec:HE,cont.lim.}.
A more general subtraction given in Eq.~(\ref{eq:HEsub,cont,def,gen}), based on
a general linear combination of the static screening correlators with coefficients 
given by $\boldsymbol\alpha$, is also possible.
The results obtained in this way have the additional index $\boldsymbol\alpha$,
which we also omitted in previous sections but reinstate in the following discussion.

\begin{figure}[t]
\begin{center}
\includegraphics[scale=0.80]{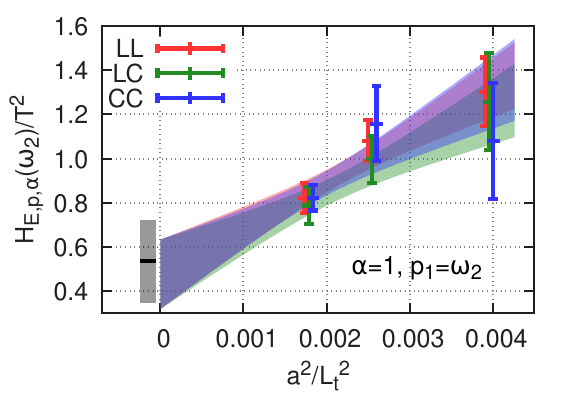}
\includegraphics[scale=0.80]{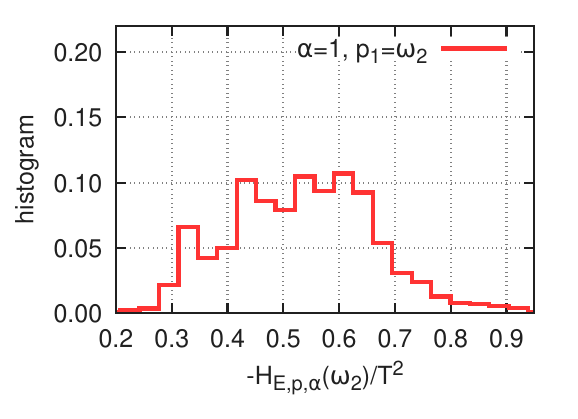}
\caption{
{\bf Left:} Representative continuum extrapolation for $H_E(\om_2)/T^2$ using
the standard subtraction (i) with $\alpha=1$ and $p_1=\om_2$ detailed in Sec.~\ref{sec:alter_sub,HE2}.
{\bf Right:} Histogram of the Akaike-weights plotted against the continuum extrapolated
results for $H_{E,p,\alpha}(\om_2)/T^2$ with $\alpha=1$ and $p_1=\om_2$.
The exact form of the integrand is in Eq.~(\ref{eq:altsub,intnd}).
}
\label{fig:HEn2,stdsub,contlim}
\end{center}
\end{figure}

We note that on a finite lattice, the values of $H_{E,{\bf p},\boldsymbol\alpha}(\om_n)$ 
with different $p_i$ and $\alpha_i$ values could differ from each other, 
but the result for $H_E(\om_n)$ in the continuum limit estimated using 
$H_{E,{\bf p},\boldsymbol\alpha}(\om_n)$ with different $p_i$ and $\alpha_i$ 
values should be the same. One can therefore explore various
choices to reduce the lattice artefacts of the continuum extrapolation.

The integrands using the general subtractions of Eq.~(\ref{eq:HEsub,cont,def,gen})
show very different behavior compared to the integrand obtained by using 
the standard subtraction with $p_1=\om_2$ and $\alpha_1=1$ ($\alpha_{i>1}=0$).
When modelling the tail of the integrands, a slightly extended version of
the modelling procedure of Sec.~\ref{sec:tail} was used due to the different
noise level of the screening correlators in the different sectors.
We discuss these procedures in more detail in App.~\ref{app:alter_sub,HE2}.
We restrict our examination to choices which are a linear combination of the static correlator with zero and non-zero momentum and with the following relation between the coefficients defined in Eq.~\eqref{eq:HEsub,cont,def,gen}
\begin{alignat}{2}
\alpha_1 &= \alpha^2, &&p_1>0\\
\alpha_2 &= 1-\alpha^2, \qquad &&p_2=0
\label{eq:altsub,intnd}
\end{alignat}
In App.~\ref{app:alter_sub,HE2}, we explore a wider range of $\alpha$ values, but here
we provide the results for only three choices of parameters:
\begin{enumerate}
    \item[(i)] the standard subtraction with $\alpha=1$ and $p_1=\om_2$
    \item[(ii)] a subtraction with $\alpha=2$ and $p_1=\omega_1$ which has the largest cancellation when Taylor-expanding in $x_3$
		among the various terms in Eq.~(\ref{eq:altsub,intnd}) at short distances
    \item[(iii)] a subtraction with $\alpha=3.5$ and $p_1=\omega_1$ which has the smallest observed slope in the continuum extrapolation
		among the investigated values of $\alpha$ (see Tab.~\ref{tab:HE2,fitI} of App.~\ref{app:alter_sub,HE2})
\end{enumerate}
We note that Eq.~\ref{eq:altsub,intnd} uses the continuum notation.

\begin{figure}[t]
\begin{center}
\includegraphics[scale=0.80]{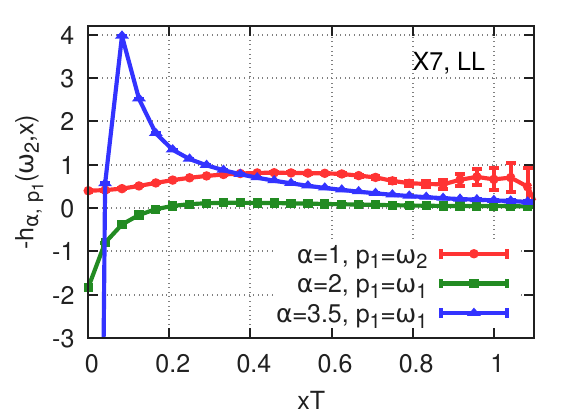}
\caption{
    Comparison of the integrands obtained by applying the standard subtraction (i) (red),
    and more general subtractions (ii) and (iii) of the form given in Eq.~(\ref{eq:altsub,intnd})
with $\alpha=2$, $p_1=\om_1$ as well as $\alpha=3.5$, $p_1=\om_1$ plotted
with green and blue colors, respectively.
The latter curve starts at around $-40$, which is not included on the plot.
}
\label{fig:HEn2,altsub,intnd}
\end{center}
\end{figure}

First, we discuss the simplest of these choices, the standard subtraction (i).
For this case, the modelling of the static screening correlator at $\om_2$
is more challenging than it was in the first Matsubara sector, because of 
the worse signal-to-noise ratio above $xT \sim 0.8$.
The continuum extrapolation has a moderate but significant slope (left panel of Fig.~\ref{fig:HEn2,stdsub,contlim}),
and the obtained continuum limit results scatter in a wide range 
as the histogram in the right panel of Fig.~\ref{fig:HEn2,stdsub,contlim} shows.
This is primarily due to the uncertainty in the modelling of the static
screening correlator at $p_1=\om_2$.
The obtained central value --- although having a large systematic error ---
is in slight contradiction with the physical expectation, namely that 
$H_E(\om_n) < H_E(\om_r)$ if $\om_n > \om_r$, which is a consequence
of the positivity of the transverse channel spectral function, $\sigma(\om)$.
This ordering of $H_E$s in various Matsubara sectors can be deduced
from Eq.~(\ref{eq:HEn-HEr,disprel}).

\begin{figure}[t]
\begin{center}
\includegraphics[scale=0.80]{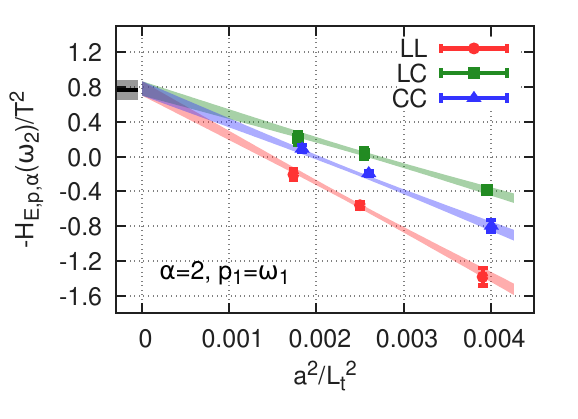}
\includegraphics[scale=0.80]{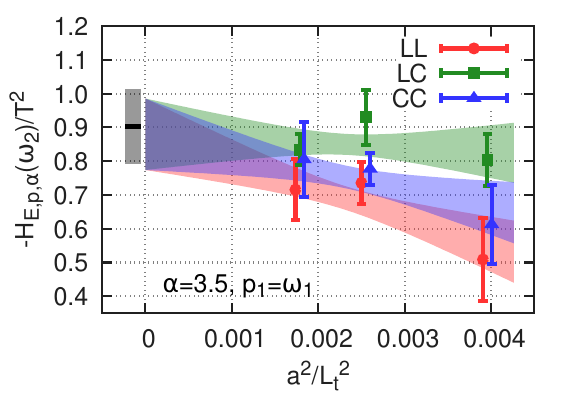}
\caption{
Representative continuum extrapolations for $H_E(\om_2)/T^2$ using
a subtraction (ii) with $\alpha=2$ and $p_1=\om_1$ ({\bf left panel})
or with choice (iii) $\alpha=3.5$ and $p_1=\om_1$ ({\bf right panel}).
The exact form of the integrand is in Eq.~(\ref{eq:altsub,intnd}).
}
\label{fig:HEn2,altsub,contlim}
\end{center}
\end{figure}

The subtractions (ii) and (iii) involving the static screening correlator at momentum $p=\om_1$ and at 
vanishing momentum is advantageous over the one at $\om_2$ primarily due to three reasons.
First, the data has better precision because it does not involve the static 
screening correlator at spatial momentum $p=\om_2$, therefore one can start the modelling 
from a later point.
The details of this modified modelling prescription are discussed in App.~\ref{app:alter_sub,HE2}.
Second, the plateau region is more clearly pronounced than in the case of the
static screening correlator in the second Matsubara sector, which results in
smaller systematic errors.
Furthermore, due to the freedom of choosing $\alpha$, one can construct integrands
which behave better, which are less weighted towards the noisy long-distance tail or have smaller cutoff effects.

As mentioned earlier, the integrand obtained using more general subtractions
can be quite different from what we obtain using the standard subtraction.
In the continuum, both $G_{\rm st}^{\rm T}(\omega_n,x_3)$ and $G_{\rm ns}^{\rm T}(\omega_n,x_3)$ are expected to have the leading
singular behaviour
\begin{equation}\label{eq:G_smallx3}
  G_{\rm st}^{\rm T}(\omega_n,x_3),~ G_{\rm ns}^{\rm T}(\omega_n,x_3)\stackrel{x_3\to0}{\sim}
  \Big(\omega_n^2 - \frac{\partial^2}{\partial x_3^2}\Big)\delta(x_3),
\end{equation}
with the same prefactor\footnote{Indeed, this comes
from the transverse tensor structure $(\partial_\mu \partial_\nu - \delta_{\mu\nu} \triangle)\delta(x)$
in the position-space vacuum correlator $\langle j_\mu(x) j_\nu(0)\rangle$.}.
This explains in particular the observed smoothness of the integrand (\ref{eq:altsub,intnd}) for our standard choice.
For subtraction (ii), the integrand would be proportional to $|x_3|$ at small $|x_3|$ in the vacuum;
therefore, the thermal integrand remains finite in the continuum at $x_3= 0$ (the difference of the static thermal and the vacuum correlator
has been investigated in~\cite{Ce:2021xgd}).
For all other values of $\alpha$, the integrand with $p_1=\omega_1$ has a singular behaviour around $x_3=0$, even though
the integral has a well-defined continuum limit.

Fig.~\ref{fig:HEn2,altsub,intnd} shows this for the second Matsubara sector,
where one can see that $-h_{\alpha=1,p_1=\om_2}(\om_2,x)$, --- i.e.\ the integrand using
the standard subtraction --- starts with a finite, positive value
at $x=0$ and after having a modest peak it decays to zero, being quite noisy
above $xT \sim 0.8$.
The integrands using alternative subtractions also go to zero at large 
distances as earlier but the decay is faster.
At $p_1=\omega_1$, the choice $\alpha=2$, results in a smooth integrand, while the one with
$\alpha=3.5$, has a more singular behaviour in the vicinity of $x_3 = 0$:
the latter integrand starts at $x/a=0$ with a value of about $-40$ for the finest ensemble
and the LL discretisation, then it changes sign at $x/a=1$ and has a peak at $x/a=2$,
after which it decays to zero.
Thus, there is a large cancellation among the point at $x_3\lesssim 1/(4T)$.
We use the trapezoid integration rule also at short distances, as we
have done earlier.

The resulting continuum extrapolations using choices (ii) and (iii) are shown in Fig.~\ref{fig:HEn2,altsub,contlim},
left and right panel, respectively.
As one can observe in Fig.~\ref{fig:HEn2,altsub,contlim}, the cutoff effects
can be markedly different using different subtractions.
For instance, the subtraction (ii) leads to a huge cutoff effect
(left panel of Fig.~\ref{fig:HEn2,altsub,contlim}), the results at the coarsest 
ensemble even have a different sign than the continuum estimate.
On the contrary, the subtraction (iii) has a very flat 
continuum extrapolation (right panel of Fig.~\ref{fig:HEn2,altsub,contlim}).
The slopes of the continuum extrapolations are also listed in Table~\ref{tab:HE2,fitI}
of App.~\ref{app:alter_sub,HE2} for these subtraction types.

The continuum estimates we obtained by applying the different subtractions
of the type given in Eq.~(\ref{eq:HEsub,cont,def,gen}) are the following:
\begin{alignat}{2}
    \textrm{(i)}&\qquad H_{E,p=\om_2,\alpha=1}(\om_2)/T^2 &&= -0.54(13)_{\rm{stat}}(13)_{\rm{sys}}, \\
    \textrm{(ii)}&\qquad H_{E,p=\om_1,\alpha=2}(\om_2)/T^2 &&= -0.76(9)_{\rm{stat}}(3)_{\rm{sys}}, \\
    \textrm{(iii)}&\qquad H_{E,p=\om_1,\alpha=3.5}(\om_2)/T^2 &&= -0.90(10)_{\rm{stat}}(4)_{\rm{sys}}. 
\end{alignat}
For further details on the modelling and the other parameters that were used
to obtain these continuum estimates, we refer to App.~\ref{app:alter_sub,HE2}.
In addition, in App.~\ref{app:alter_sub,HE2}, we investigate a broader set of $\alpha$
values, from which we conclude that the continuum result obtained 
by having a flat continuum extrapolation with $\alpha=3.5$ is a result that
is in good agreement with more or less all the other results 
(c.f. Fig.~\ref{fig:HE2,fitI} and also Fig.~\ref{fig:HE2,fitII}).
Therefore, we choose this value,
\beq
	H_E(\om_2)/T^2 = -0.90(10)_{\rm{stat}}(4)_{\rm{sys}}
\eeq
as our final continuum estimate in the second Matsubara sector.

\subsection{Direct evaluation of the difference $H_E(\om_2)-H_E(\om_1)$}
\label{sec:HE2-HE1}

Using Eq.~\ref{eq:HEn-HEr,cont,def,gen}, we can directly evaluate the difference 
of $H_E$s obtained in different Matsubara sectors, or more generally the difference 
$H_E(\om_r)-\varepsilon H_E(\om_n)$.
The results obtained in this way can be compared to the results obtained by
calculating $H_E(\om_r)$ and $H_E(\om_n)$ separately and thereby serve as
a useful crosscheck of those results.
Additionally, the difference is an interesting quantity for its own sake.
For instance, the choice $\varepsilon=1$ probes an integrand, which is non-singular 
and is only sensitive to photons at nonzero frequencies (see Fig.~\ref{fig:kernel,cont}).

We calculated the linear combination $H_E(\om_2)-\varepsilon H_E(\om_1)$ in several ways
and summarized the results in Table~\ref{tab:HE2,fitII} of App.~\ref{app:alter_sub,HE2}.
Among the results listed in Table~\ref{tab:HE2,fitII}, we highlight here the ones
which have the flattest continuum extrapolations.
These are the subtractions that have
\beq
    \epsilon = 1,\qquad\alpha_1=11.25,\qquad\alpha_2=-11.25,
\eeq
and
\beq
    \epsilon = 14,\qquad \alpha_1 = 0,\qquad \alpha_2=-13.
\eeq
In both cases, the momentum of the subtracted non-static screening correlator
is $\om_1$ and the momenta of the static screening correlators 
are $p_1=\om_1$ and $p_2=0$, for the correlators multiplied 
by $\alpha_1$ and $\alpha_2$, respectively.
Thus, the integrand in the continuum has the exact form of
\beq
	G_{\rm{ns}}^\sT(\om_2, x_3) \eexp^{-\om_2 x_3} 
		  - \varepsilon\, G_{\rm{ns}}^\sT(\om_1, x_3) \eexp^{-\om_1 x_3} 
		  - \alpha_1\, G_{\rm{st}}^\sT(\om_1, x_3) \eexp^{-\om_1 x_3}
		  - \alpha_2\, G_{\rm{st}}^\sT(0, x_3).
\label{eq:altsub,intnd,a0}
\eeq

The results for $\varepsilon=1$ correspond to
forming the difference of the integrand for $H_E(\om_2)$ that has a flattest observed 
continuum extrapolation (choice (iii) of the previous section, 
see Eq.~\ref{eq:altsub,intnd}) and the integrand for $H_E(\om_1)$ having
a standard subtraction term (Eq.~\ref{eq:HEsub,cont,def} with $n=1$ and $p=\om_1$).
Therefore, it is nice to observe that the difference obtained in the continuum limit
\beq\label{eq:HE2-HE1}
	\frac{1}{T^2}\Big[ H_E(\om_2) - H_E(\om_1) \Big]_{\boldsymbol{p}=(\om_1,0),\,\boldsymbol{\alpha}=(11.25,-11.25)} = -0.23(12)_{\rm{stat}}(4)_{\rm{sys}}
\eeq
is in complete agreement with the difference formed by using the continuum
estimates for $H_E(\om_2)$ and $H_E(\om_1)$, separately.

For the other set of parameters with $\epsilon=14$ for which we also observe a mild continuum extrapolation, we obtain
\beq
	\frac{1}{T^2}\Big[ H_E(\om_2) - \varepsilon H_E(\om_1) \Big]_{\boldsymbol{p}=(\om_1,0),\,\boldsymbol{\alpha}=(0,-13)} = 8.44(11)_{\rm{stat}}(5)_{\rm{sys}}.
\eeq
In order to put this into context, we calculated $H_E(\om_2)/T^2$ from this result
by adding 14 times the continuum estimate of $H_E(\om_1)/T^2$ in a correlated way.
We obtained $H_E(\om_2)/T^2 = -0.94(12)_{\rm{stat}}(5)_{\rm{sys}}$,
which is also in good agreement with our final estimate.

\section{Comparisons}
\label{sec:comp}

In this section, we compare our findings to results from analytic approaches.
We start with the integrand for $H_E(\om_1)$, which is shown in
Fig.~\ref{fig:HE,intnd_X7_vs_free,AMY}.
The calculation of the integrand of the free continuum result requires special
care at very short distances, below $xT \sim 0.1$.
This is discussed in App.~\ref{app:free_th}.
By looking at Fig.~\ref{fig:HE,intnd_X7_vs_free,AMY}, 
we can observe that the continuum free theory result has very different 
characteristics compared to the lattice QCD result.
Although both start with a finite value at $x=0$, the free theory result decays
faster and even goes to slightly negative values above $xT \approx 1$.

\begin{figure}[t]
\begin{center}
\includegraphics[scale=0.80]{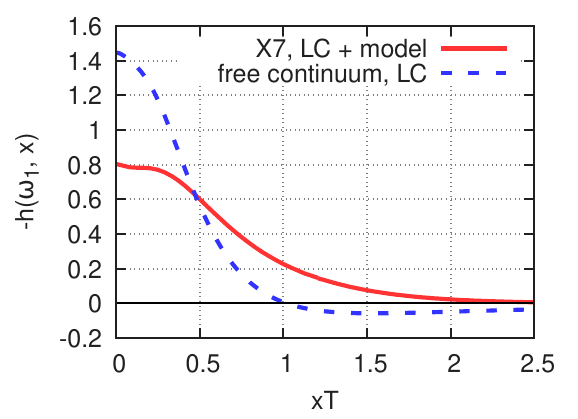}
\caption{
  The integrand for determining $H_E(\om_1)$ in the free theory (dashed)
and using lattice QCD data on our finest ensemble X7, completed with a model
based on single-state fits for long distances using
Eqs.~(\ref{eq:HEsub,lat,intnd0}),(\ref{eq:h,split}),(\ref{eq:h_ans})
(solid line).
Only the central values are shown.
}
\label{fig:HE,intnd_X7_vs_free,AMY}
\end{center}
\end{figure}

Secondly, we summarized the comparison of the non-static screening masses determined 
on the lattice and the weak-coupling results in Fig.~\ref{fig:comp,E,n1}.
We refer to App.~\ref{app:weak-cth} as well as to Ref.~\cite{Brandt:2014uda} 
for more detailed information about the results.
As Fig.~\ref{fig:comp,E,n1} shows, the lattice result for the non-static screening 
mass in the $n=1$ sector lies between the NLO and the EQCD result.
We conclude that the weak-coupling prediction is fairly successful, at the $4\%$ level, provided
at least the next-to-leading order interquark potential is used.
The weak-coupling results in Fig.~\ref{fig:comp,E,n1} are based on the value $\alpha_s=0.25$,
which will be our default value of the gauge coupling in the following.

\begin{figure}[t]
\begin{center}
\includegraphics[scale=0.77]{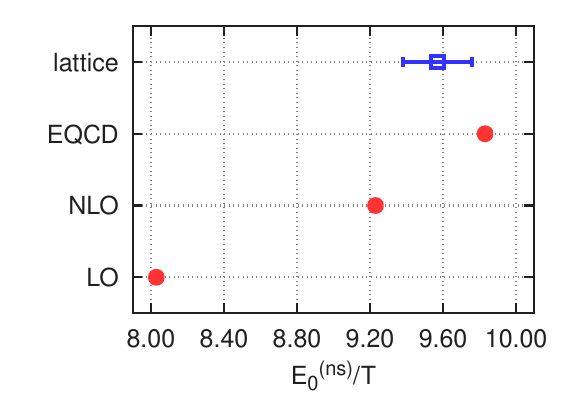}
\caption{
Ground-state non-static screening masses in the $n=1$ Matsubara sector
determined on the lattice and by using weak-coupling theory at leading order (LO),
with an interquark potential at next-to-leading order (NLO),
and by using the potential at $T=400$ MeV of the dimensionally
reduced effective theory for QCD, 'electrostatic QCD' (EQCD).
}
\label{fig:comp,E,n1}
\end{center}
\end{figure}

Calculating the imaginary part of the retarded correlator at lightlike kinematics
in the free massless theory with our current normalization gives $H_E(\om_n)/T^2= -N_c/6$, independently of $n$~\cite{Meyer:2018xpt}.
This corresponds to a free spectral function
\beq\label{eq:freespecfun}
\sigma^{\rm free}(\om)/\omega = \frac{N_c}{3} \pi T^2\,\delta(\om), \qquad \chi_s^{\rm free} = \frac{N_c}{3}T^2.
\eeq
Therefore~${H_E(\om_1)/\chi_s^{\rm free}=-0.500}$. 
This result can also be reproduced to good precision by integrating the
free theory integrand shown in Fig.~\ref{fig:HE,intnd_X7_vs_free,AMY},
if one uses a suitable representation (cf.\ appendix~\ref{app:free_th}).

In strongly coupled $\mathcal{N}=4$ super Yang--Mills (SYM) theory using
the AdS/CFT correspondence, one finds $H_E(\om_1)/\chi_s^{\rm free} = -0.336$~\cite{Meyer:2018xpt},
using $(\chi_s)^{\rm free} = N_c^2 T^2/4$ (see~\cite{Teaney:2006nc}, appendix~A).
This value is in fact lower than the free-theory result.
Compared in this way, the lattice result we obtained, $-0.670(6)_{\rm{stat}}(1)_{\rm{sys}}$,
is largest.

Using a different normalization, e.g.\ dividing by the static susceptibility, $\chi_s$,
of the relevant \emph{interacting} theories we arrive at the following predictions:
in the free theory, obviously the result does not change, while 
in $\mathcal{N}=4$ super Yang--Mills theory, we get $[H_E/\chi_s]^{\rm{(SYM)}} = -0.6715$~\cite{Meyer:2018xpt},
where $\chi_s = N_c^2 T^2/8$.
In Ref.~\cite{Ce:2022fot}, we determined the static susceptibility at this
temperature to be
\beq
\chi_s^{\rm{(lat)}}/T^2=0.882(11)_{\rm{stat}}(19)_{\rm{sys}}.
\eeq
Using this value, our lattice result is $[H_E/\chi_s]^{\rm{(lat)}} \simeq -0.76$.
Thus, now normalizing by the interacting $\chi_s$, the lattice result is still the largest in magnitude.
These results are illustrated in Fig.~\ref{fig:comp,HE,n1}.

In the second Matsubara sector, we have $-1.115$ for $H_E(\om_2)/T^2$ 
in strongly coupled $\mathcal{N}=4$ SYM~\cite{Meyer:2018xpt}, thus the difference $(H_E(\om_2)-H_E(\om_1))/\chi_s$ is $-0.444$,
a value larger than our lattice result by 1.7 standard deviations.
Since $H_E(\om_n)$ is constant in the free theory, the difference vanishes there.
Thus the ratio $(H_E(\om_2)-H_E(\om_1))/H_E(\om_1)$ provides good sensitivity to the shape of $\sigma(\om)$,
being 0.66 in the strongly-coupled SYM case and parametrically small in the case of a spectral function very peaked around $\omega=0$.
The lattice result, $(H_E(\om_2)-H_E(\om_1))/H_E(\om_1)= 0.34\pm0.19$, lies between these two extremes.

\subsection{Computing $H_E(\om_n)$ dispersively using the complete leading-oder $\sigma(\om)$}

Using the complete leading-order result of Arnold, Moore and Yaffe (AMY) for $\sigma(\om)$~\cite{Arnold:2001ba,Arnold:2001ms},
the difference $[H_E(\om_2)-H_E(\om_1)]$ can be evaluated straightforwardly, whereas
the quantity $H_E$ in individual Matsubara sectors can only be estimated after handling
the singular behavior of $\sigma_{\rm{AMY}}(\om)$ at small frequencies.
Indeed, by integrating the spectral function of Ref.~\cite{Arnold:2001ba,Arnold:2001ms} (with $\alpha_s=0.25$) 
multiplied by the kernel $\om_n^2/(\pi \om (\om^2+\om_1^2))$ in the range $0.2 < \om/T < 50$
where the provided parametrisation is a good approximation, we obtain 
\beq\label{eq:HE1AMY}
	\Big[H_E(\om_1)/T^2\Big]_{\rm{LO},\,0.2<\om/T<50} \approx -0.75,
\eeq
which 
 is already larger in absolute value than the result obtained on the lattice.
On the other hand, the prediction for $H_E(\om_2)-H_E(\om_1)$ using the AMY spectral function
is much less sensitive to the small-$\omega$ behaviour. 
We obtain, again with $\alpha_s=0.25$,
\beq\label{eq:HE2-HE1AMY}
	\Big[ H_E(\om_2)/T^2 - H_E(\om_1)/T^2 \Big]_{\rm{LO},\,0.2<\om/T<50} \approx -0.25.
\eeq
The comparison can be seen in Fig.~\ref{fig:comp,HE,n1}, right panel.
The leading-order prediction for this difference is well compatible with our lattice QCD result.
We also note that extending the integral to $\om=\infty$ assuming $\sigma(\om)\propto \sqrt{\om}$~\cite{Arnold:2001ms}
changes the value of Eq.\ (\ref{eq:HE2-HE1AMY}) to $-0.28$.

\begin{figure}[t]
\begin{center}
\includegraphics[scale=0.75]{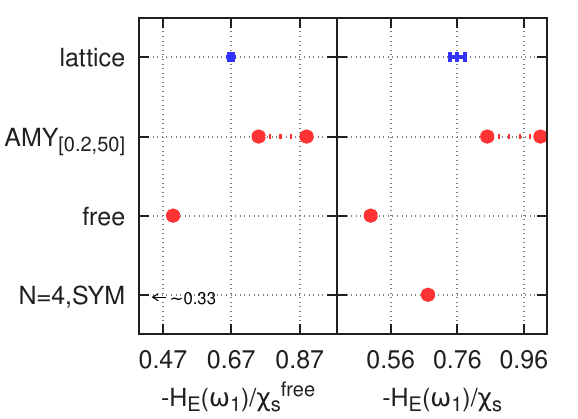}
\includegraphics[scale=0.77]{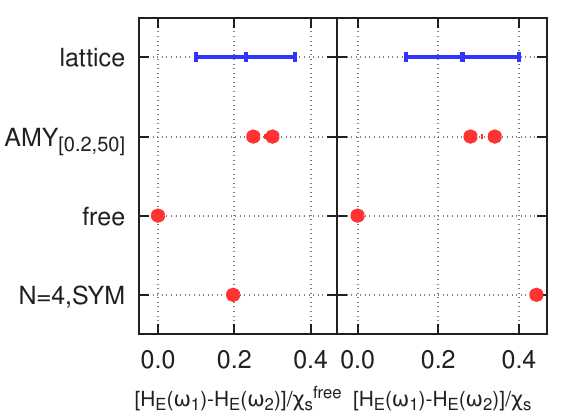}
\caption{
{\bf Left}: Comparison of results for $H_E(\om_1)$ obtained on the lattice,
in leading order of the weak-coupling expansion (with $\alpha_s=0.25$ ($\alpha_s=0.31$) 
for the left (right) point), in the free theory  and in strongly 
coupled $\mathcal{N}=4$ SYM, normalized by the static susceptibility
of the corresponding non-interacting theory (left inset),
or by the interacting static susceptibility (right inset).
We interpret the two normalizations as being identical for the ``free'' data points,
while to normalize the NLO results we used alternatively 1 and the lattice result 0.88 for $\chi_s/T^2$.
{\bf Right}: Comparison for $H_E(\om_2)-H_E(\om_1)$ with the same choices of normalization.
}
\label{fig:comp,HE,n1}
\end{center}
\end{figure}

In order to exploit our precise lattice result for $H_E(\om_1)$, we need to inspect more precisely the
region of validity of the weak-coupling spectral function.
The leading-order calculation~\cite{Arnold:2001ba,Arnold:2001ms} assumes the photon wavelength to be short compared to
the mean free path for large-angle scattering.
At the smallest frequencies at which the calculation is still valid, $\sigma(\omega)\propto 1/\sqrt{\omega}$.
Since we expect $\sigma(\omega)/(2\chi_s\omega)$ to tend to a finite value at $\om\to0$,
namely to the diffusion coefficient $D$ (see Eq.\ (\ref{eq:Dfromsigma})),
it is clear that a qualitative change in the functional form must take place below the frequency at which
the AMY calculation breaks down.
In addition, the next-to-leading order correction~\cite{Ghiglieri:2013gia}, suppressed only by one power of the strong coupling $g$,
turns out to be quite modest (about $\pm 10\%$) for $\omega \gtrsim 2\pi T$, but becomes larger (about $\pm 30\%$)
at  $\omega\lesssim \pi T$.

Thus a qualitative modification of the AMY spectral function at small frequencies is necessary
for a sensible dispersive evaluation of $H_E(\om_1)$.
In the following, we assume that $\sigma(\om)$ is given by the leading-order expression~\cite{Arnold:2001ms} 
for $\om>\om_{\rm m}$, introduce a Lorentzian ansatz for $\om<\om_{\rm m}$,
\beq\label{eq:Lorentzian}
\frac{\sigma(\om)}{2\chi_s \om} = \frac{A}{\omega^2 + \eta^2},
\eeq
and require continuity and differentiability at $\om=\om_{\rm m}$.
Depending on $\om_{\rm m}$, different values of $H_E(\om_1)$, and of $D=A/\eta^2$, are obtained.
At the high-frequency end, we extend $\sigma(\om)$ beyond $\om=50T$ assuming $\sigma(\om)\propto \sqrt{\om}$,
though this high-energy region only contributes about $-0.01$ to $H_E(\om_1)/T^2$.

The leading-order weak-coupling result for $D$ is known from Ref.~\cite{Arnold:2003zc}:
with $m_D /T  = 2.05$, which results from setting $\alpha_s=0.25$ in the leading order expression
of the Debye mass, we read off $T\cdot D  = 2.3$ from Fig.\ 1 of~\cite{Arnold:2003zc}.
It turns out that this value of the diffusion coefficient can be reproduced by choosing the matching point
at $\om_{\rm m}=0.50T$, using $\alpha_s=0.25$ as well in the weak-coupling spectral function $\sigma(\om)$~\cite{Arnold:2001ms}.
In that case, however, the dispersive integral yields $H_E(\om_1)/T^2=-0.99$, in stark disagreement with our lattice result
(\ref{eq:HE1,final,cont}). Thus this particular weak-coupling scenario is ruled out at $T=250\,$MeV by our lattice calculation.

\begin{figure}[t]
\begin{center}
\includegraphics[scale=0.77]{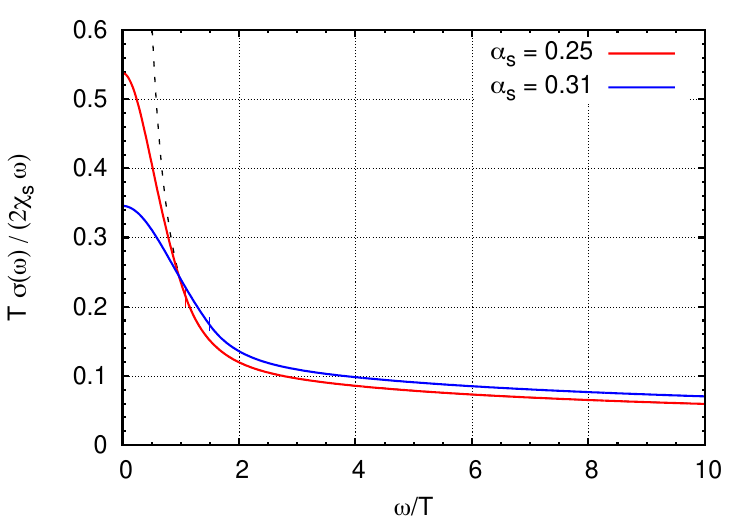}
\caption{
  Two weak-coupling based spectral functions reproducing the value (\ref{eq:HE1,final,cont}) for $H_E(\om_1)$.
  Beyond a value $\om=\om_{\rm m}$ marked by a short vertical line, the curves correspond to the complete leading-order result
  in the parametrization provided by~\cite{Arnold:2001ms}, for two values of $\alpha_s$. For $\om<\om_{\rm m}$,
  the curves correspond to the Lorentzian (\ref{eq:Lorentzian}) with parameters given in Eqs.\ (\ref{eq:D0p25}--\ref{eq:D0p31}), respectively.
  In the $\alpha_s=0.25$ case, the dashed curve shows how the parametrization of~\cite{Arnold:2001ms} extends towards smaller frequencies.
  The intercept at $\om=0$ yields the isospin diffusion coefficient, $T\cdot D$.
}
\label{fig:sf_weak}
\end{center}
\end{figure}

Instead of assuming the weak-coupling result for the diffusion coefficient,
we can attempt to estimate it by continuing
the leading-order expression~\cite{Arnold:2001ms} of $\sigma(\om)$ 
toward the soft-photon limit via the Lorentzian Eq.\ (\ref{eq:Lorentzian})
so as to reproduce our lattice result for $H_E(\om_1)$.
With $\alpha_s=0.25$, this condition yields 
\beq\label{eq:D0p25}
\alpha_s=0.25: \quad 
\om_{\rm m}= 1.08\,T, \qquad A=0.42\,T, \qquad \eta=0.88\,T, \qquad T\cdot D= 0.54.
\eeq
Increasing the targeted $|H_E(\om_1)|/T^2$ by one standard deviation (0.006) only results in a modest change
in the estimated diffusion coefficient to $T\cdot D= 0.56$.
Thus the estimate of the diffusion coefficient obtained in this way is much lower than the
weak-coupling result~\cite{Arnold:2003zc}, but still more than three times larger than the AdS/CFT
value of $D=(2\pi T)^{-1}$.
Repeating the procedure above with a larger value of the coupling, we obtain
\beq \label{eq:D0p31}
\alpha_s=0.31: \quad 
\om_{\rm m}= 1.50\,T, \qquad A=0.78\,T, \qquad \eta=1.50\,T, \qquad T\cdot D= 0.35.
\eeq
In summary, under the assumptions made on the spectral function, we obtain a range of diffusion coefficients
given by Eqs.\ (\ref{eq:D0p25}--\ref{eq:D0p31}).
The corresponding spectral functions are illustrated in Fig.~\ref{fig:sf_weak}.

It is worth recalling that in the non-interacting theory, the spectral function has the form $\om\delta(\om)$
(see Eq.\ (\ref{eq:freespecfun})).
In the limit of vanishing coupling, one thus expects the Lorentzian to turn into a delta function
($\eta\to0$ at fixed $A/\eta$),
and the spectral function to vanish roughly proportionally to $\alpha_s$ for $\om\gg \eta$.
We define the area under the peak around $\om=0$ as follows,
\beq\label{eq:area0}
S_0 \equiv \int_{-\Omega}^\Omega \frac{d\om}{\pi}\; \frac{\sigma(\om)}{2\chi_s \om} ,
\eeq
where $\Omega\gg \eta$ is a UV-cutoff.
On the one hand, we get with our Lorentzian ansatz  $S_0 = {A}/{\eta}$.
In the free theory, one obtains $S_0 = 1/2$. Thus it is remarkable that
the ratios $A/\eta$ in Eqs.\ (\ref{eq:D0p25}--\ref{eq:D0p31}) are close to $1/2$:
the obtained values of the parameters $(A,\eta)$ are plausible in this respect,
since our analysis is based on an  ansatz for the spectral function valid at weak coupling
and $S_0$ has a weak-coupling expansion of which $S_0=1/2$ is the leading term.

\section{Conclusions and outlook}
\label{sec:conclusions}

The thermal photon emissivity of the quark-gluon plasma is determined
to all orders in the strong coupling by the transverse channel spectral function
of electromagnetic current-current correlators evaluated at lightlike kinematics.
This real-time observable is not accessible directly on a Euclidean lattice.
In this work, we demonstrated that it is nonetheless possible to directly evaluate
an observable in lattice QCD that is related to the aforementioned
spectral function via a dispersion relation.
Indeed, the computed observable --- a Euclidean screening correlator at imaginary spatial 
momentum --- has an integral representation in terms of a product of the spectral 
function multiplied by a Lorentzian kernel (see Eq.\ \ref{eq:HE,Q^2=0}).
However, the naive lattice estimator of this observable is afflicted by a large cutoff effect
arising from the breaking of Lorentz invariance on the lattice. 
We successfully addressed this technical difficulty
by subtracting screening correlators at different momenta, whose
contribution vanishes in the continuum.
We investigated various subtractions which lead to different scaling behaviors
towards the continuum limit.

By analyzing the long-distance behavior of the screening correlators,
we also determined the screening masses (see Eqs.\ (\ref{eq:m0st},\ref{eq:E0st,n1},\ref{eq:E0ns,n1})),
which we used to model the position-space correlators
at long distances and improve the precision of the imaginary-momentum correlators.
In order to probe the virtuality dependence of the latter around the $Q^2=0$ point, 
we constructed a suitable lattice representation of its $Q^2$-derivative 
and evaluated it in the first Matsubara sector.
We used two-flavors of O($a$)-improved Wilson fermions at a temperature
of about 250\;MeV with an \emph{in vacuo} pion mass of 270\;MeV and performed continuum 
extrapolations of all our observables.

Our final continuum result for $H_E(\om_1)/T^2$, see Eq.\ (\ref{eq:HE1,final,cont}), has a precision of around 1\%,
suitable for a comparison with other approaches.
We confronted our results to estimates using the free theory, strongly-coupled
$\mathcal{N}=4$ SYM and the full leading-order result of Arnold, Moore and Yaffe (AMY)~\cite{Arnold:2001ba,Arnold:2001ms}.
We found that our result for $|H_E(\om_1)|$ is smaller than the prediction 
obtained by using the spectral function of Ref.~\cite{Arnold:2001ba,Arnold:2001ms},
although our result for $H_E(\omega_2) - H_E(\omega_1)$, given in Eq.\ (\ref{eq:HE2-HE1}), is comparable.
This is illustrated in Fig.~\ref{fig:comp,HE,n1}.
Since the integration kernel is much more sensitive to the low-frequency behavior
of the spectral function in the case of $H_E(\om_n)$ than in the case of the 
difference (c.f. Fig.~\ref{fig:kernel,cont}), our result suggests that 
the weak-coupling result for the spectral function is
overestimated in this low-frequency region. 
Assuming the AMY spectral function to hold above a certain frequency $\om_{\rm m}$,
and using a Lorentzian transport peak below that frequency that matches on smoothly at $\om_{\rm m}$,
we arrive at estimates of the isospin diffusion coefficient $T\cdot D$ in the range 0.35 to 0.54
(see Eqs.\ \ref{eq:D0p25}--\ref{eq:D0p31}) by requiring that our lattice result for $H_E(\om_1)$ be reproduced.
This range of values for $T\cdot D$ is in line with previous lattice estimates based on dispersion relations
at fixed spatial momentum~\cite{Brandt:2012jc,Amato:2013naa,Aarts:2014nba,Brandt:2015aqk,Ghiglieri:2016tvj,Ding:2016hua,Astrakhantsev:2019zkr},
though the central values of most calculations with dynamical quarks lie below the value
of 0.3 at temperatures around 250\,MeV~\cite{Aarts:2020dda}.

It is also interesting to compare our results with our two previous studies of the photon emissivity
that employed the same gauge ensembles as the present calculation but were
based on the dispersion relation at fixed spatial momentum~\cite{Ce:2021xgd,Ce:2022fot}. Addressing the inverse problem
with physically motivated ans\"atze for the spectral functions, we concluded~\cite{Ce:2022fot}
(particularly for $\omega\geq\pi T$) that the lattice results 
were consistent with the weak-coupling prediction, but could also accomodate a rate 2.5 times larger.
In other words, most of the solutions for the spectral function describing the lattice data yielded a photon rate at least as large
as the weak-coupling prediction.
Given that the weak-coupling spectral function~\cite{Arnold:2001ms}
results in a larger value for $|H_E(\omega_1)|$ than our lattice result,  it seems 
most likely that this excess is due to an overestimated soft-photon emissivity in the weak-coupling calculation.

We have seen that computing $H_E(\omega_2)$ by standard lattice methods is already a lot more challenging than for $H_E(\omega_1)$
due to the large statistical errors on the position-space integrand.
Thus it would be interesting to investigate noise-reduction methods similar to those used in calculations of the hadronic vacuum polarization.
Improving on our determination of $H_E(\omega_2) - H_E(\omega_1)$ would go a long way to ascertain that the emission rate of hard photons
$(\om\gtrsim \pi T)$ follows the weak-coupling prediction at the temperatures reached in heavy-ion collisions. This appears to
be an achievable goal in the near future.

A second difficulty that occurred is the increase in discretization
errors for increasing $\om_n$. Perhaps some improvements in the
discretization are still possible here, but it is clear that very fine
lattices are necessary in order to determine $H_E(\om_3)$ and beyond.
While the moments $H_E(\om_n)$ provide valuable non-perturbative
constraints on the spectral function, a numerically ill-posed inverse
problem resurfaces if one has the ambition of determining
$\sigma(\omega)$ itself from a (necessarily) finite collection of its
moments.
Instead, our immediate plan is to compute the first moment of $\sigma(\omega)$ across the phase crossover with physical quark masses in order
to test the models of $\sigma(\omega)$ used in hydrodynamics-based calculations of photon spectra in heavy-ion collisions.

\section{Acknowledgements}

This work was supported by the European Research Council (ERC) under the European
Union’s Horizon 2020 research and innovation program through Grant Agreement
No.\ 771971-SIMDAMA, as well as by the Deutsche Forschungsgemeinschaft 
(DFG, German Research Foundation) through the Cluster of Excellence “Precision Physics,
Fundamental Interactions and Structure of Matter” (PRISMA+ EXC 2118/1) funded by
the DFG within the German Excellence strategy (Project ID 39083149).
The research of M.C. is funded through the MUR program for young researchers ``Rita Levi Montalcini''.
The generation of gauge configurations as well as the computation of correlators was
performed on the Clover and Himster2 platforms at Helmholtz-Institut Mainz and on Mogon II
at Johannes Gutenberg University Mainz.
We have also benefited from computing resources at Forschungszentrum J\"ulich allocated
under NIC project HMZ21.
For generating the configurations and performing measurements, we used the openQCD~\cite{Luscher:2012av}
as well as the QDP++ packages~\cite{Edwards:2004sx}, respectively.


\appendix
\section{Elimination of outliers}
\label{app:outliers}

\begin{figure}[t]
\begin{center}
\includegraphics[scale=0.72]{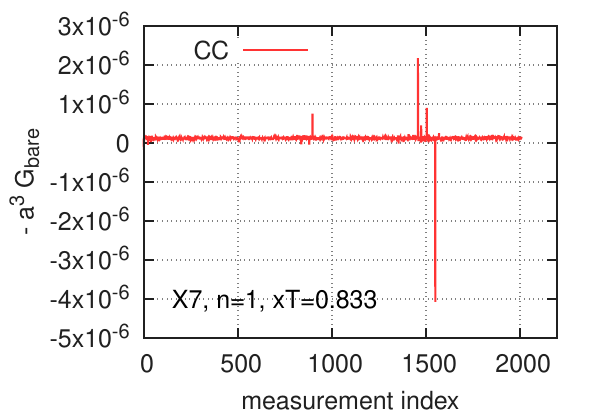}
\includegraphics[scale=0.75]{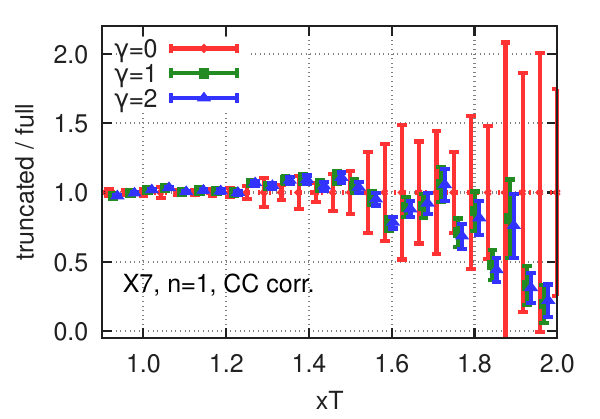}
\caption{{\bf Left}: Measurement history of the non-static screening correlator at $xT=0.833$.
The outliers are shown up as spikes in the data.
{\bf Right}: Truncation stability for the non-static screening correlator.
The data points corresponding to the trimmed data have been shifted slightly
to the right to improve visibility.}
\label{fig:outliers&truncation}
\end{center}
\end{figure}

In this Appendix, we describe our procedure to deal with certain
measurements that deviate by several standard deviations from the
mean value calculated using the data.
We identify these exceptional measurements as outliers, see the left panel
of Fig.~\ref{fig:outliers&truncation} and we found, they occur more
frequently at large Euclidean separations.
These outliers increased the statistical error and also modified the mean
to some extent, see Fig.~\ref{fig:outliers&truncation}, right panel.
We eliminated them by using robust statistics~\cite{Huber:2009snd}.

In our procedure, we first prepared a distribution of results at each
Euclidean distance, then removed the data points belonging to the lower
and upper $\gamma$\% of that distribution.
Varying $\gamma$ in the interval 0.5--4, we made cuts and found that
the error estimation as well as the calculation of the mean is more stable
this way.
When we detected only less than 10 datapoints being outside five times the
interquantile range from the mean, we applied only a trimming with $\gamma=0.5$.
For other distances we applied $\gamma=1$ in our final analysis.

We show an example in the case of the conserved-conserved correlator on our finest
ensemble, X7, in the right panel of Fig.~\ref{fig:outliers&truncation}.
At short distances of the correlator, this approach did not influence the
results, because outliers occured there only very rarely.
At intermediate distances, i.e.\ around $xT\sim 0.7$--$1.3$, the effect of this
method was again not significant.
At large distances, however, the errors reduced by a factor of around 2--6
when omitting the tails of the distributions.
Although one has to be careful when discarding certain measurements,
we believe that our procedure removing outliers sometimes more than 10 standard
deviations off from the mean at large distances should not influence 
the validity of the extracted physical results.

\section{Alternative subtractions for determining $H_E(\om_1)$}
\label{app:alter_sub,HE1}

In this appendix, we discuss the determination of $H_E(\om_1)$ using 
a more general class of subtractions to tame the short-distance cutoff effects.

Besides the standard subtraction in the first Matsubara sector 
(Eq.~(\ref{eq:HEsub,cont,def}) with $p=\om_1$),
we calculate $H_E(\om_1)/T^2$ by integrating the integrand formed 
by subtracting the completely static screening correlator ($n=0$) 
from the non-static screening correlator at the $n=1$ Matsubara sector.
This latter is denoted by $H_{E,p=0}(\om_1)/T^2$, while the results
obtained using the standard subtraction are denoted by $H_{E,p=\om_1}(\om_1)/T^2$
in this appendix.
We note, that in the main text, we left the lower subscript $p$, because we
used the standard subtraction when discussing $H_E$ in the first Matsubara sector
in~Secs.~\ref{sec:HE,intnd},\ref{sec:tail}~and~\ref{sec:HE,cont.lim.}.

The integrand obtained by subtracting the completely static screening correlator ($p=0$)
is quite different than it was in the case of subtracting the static correlator at $p=\om_1$.
At larges distances it goes to zero as earlier, but with different sign.
It receives a huge contribution from very short distances, $xT \sim 0$,
since it starts at $x/a=0$ with a large value, then it changes sign at $x/a=1$.

We emphasize here that on a finite lattice, the values of $H_{E,p}(\om_n)$ with different
$p$ differ from each other, but the result for $H_E(\om_n)$ in the continuum limit 
estimated using $H_{E,p}(\om_n)$ with different $p$ values should agree.
In Fig.~\ref{fig:HE,contlim,mom0}, we compare the continuum extrapolations
using $p=\om_1$ --- which was our standard choice in previous sections ---
to using $p=0$, i.e.\ when we subtract the completely static, zero-momentum
screening correlator when forming the integrand in Eq.~(\ref{eq:HEsub,cont,def}).

As one can observe on Fig.~\ref{fig:HE,contlim,mom0}, the continuum extrapolation
is much steeper in the case of $H_{E,p}(\om_1)$ with $p=0$, although the
linear scaling in $a^2$ persists.
Repeating a similar continuum procedure for $H_{E,p}(\om_1)$ with $p=0$ as
was discussed for $H_{E,p}(\om_1)$ with $p=\om_1$ in Sec.~\ref{sec:HE,cont.lim.},
we found quantitatively similar fit qualities.
Using $p=0$ in Eq.~(\ref{eq:HEsub,cont,def}), the estimate 
for $H_E(\om_1)/T^2$ in the continuum is
\beq
	\lim_{a^2 \to 0} H_{E,p=0}(\om_1;a)/T^2 = -0.653(6)_{\rm{stat}}(2)_{\rm{sys}},
\eeq
which is 1.9 standard deviations smaller than the continuum
results obtained using the data with $p=\om_1$, c.f. Eq.~(\ref{eq:HE1,final,cont}).

This discrepancy can be either attributed to the breakdown of the trapezoid 
integration around $xT \sim 0$ or to the steeper continuum extrapolation
and the presence of higher order lattice artifacts which we could not resolve 
at the current precision.
On the other hand, this agreement to 1.9\,$\sigma$ is quite remarkable
in view of the different magnitude of the $a^2$ coefficients 
in the continuum extrapolation, which are more than an order of magnitude
larger in absolute value when using $p=0$ instead of $p=\om_1$.

\begin{figure}[t]
\begin{center}
\includegraphics[scale=0.80]{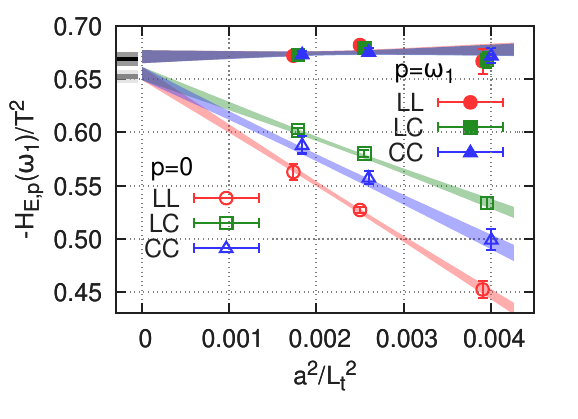}
\caption{
Comparison of representative continuum extrapolations for $H_E(\om_1)/T^2$ using
$H_{E,p=\om_1}(\om_1)/T^2$ or $H_{E,p=0}(\om_1)/T^2$ with filled
and open symbols, respectively.
}
\label{fig:HE,contlim,mom0}
\end{center}
\end{figure}

\section{More details on the alternative subtractions for determining $H_E(\om_2)$}
\label{app:alter_sub,HE2}

In this appendix, we discuss more details about the more general alternative
subtractions of which we presented a few results in Sec.~\ref{sec:alter_sub,HE2}.

\begin{figure}[t]
\begin{center}
\includegraphics[scale=0.80]{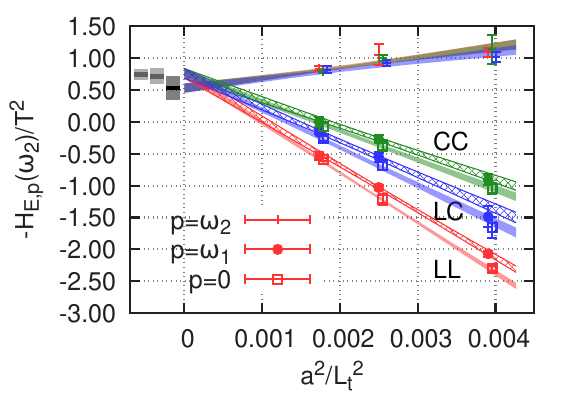}
\caption{
Comparison of representative continuum extrapolations for $H_E(\om_2)/T^2$
using $\alpha=1$ and $p_1=\om_2$, $p_1=\om_1$ or $p_1=0$.
Different colors correspond to different discretizations
(local-local (LL): red, local-conserved (LC): blue, conserved-conserved (CC): green).
Note the gigantic lattice artifacts when subtracting the static screening correlator
with $p_1=\om_1$ or $p_1=0$.
}
\label{fig:HEn2,altsubI,contlim}
\end{center}
\end{figure}

When subtracting the static screening correlator at $p=\om_2$,
we applied the same procedure as was discussed for $H_E(\om_1)/T^2$
in Sec.~\ref{sec:tail}.
Since the data in the second Matsubara sector is noisier, we had to 
start the modelling from an earlier distance, therefore we applied 
$x_{\rm{w,ns}} T = 0.6, 0.7$.

When subtracting the static screening correlator at $p=\om_1$ or at $p=0$,
we slightly modified the procedure of modelling the integrand.
We divided the integration interval into three parts.
At short distances, both the non-static and static screening correlators have
good signal-to-noise ratios, therefore we applied no modelling there.
At the intermediate interval, the non-static screening correlator
is already modelled but the static screening correlator is not.
In this interval, we calculate the integral using the trapezoidal rule
applied for the integrand formed by evaluating the single-state fit used for 
modelling the non-static screening correlator at the lattice points and 
subtracting the lattice data for the static screening correlators.
Finally, in the interval for large distances, we modelled the static
screening correlators as well and applied Eqs.~(\ref{eq:h_ans})~and~(\ref{eq:int,exptanh}).
We start applying modelling by single-state fits using the smooth step function
of Eq.~(\ref{eq:step_fun}) from $x_{w,\rm{ns}} T = 0.6, 0.7$ in the
case of the non-static screening correlator in the second Matsubara sector
and from $x_{w,\rm{st}} T=1.0,1.1,1.2,1.3$ in the case of the static screening 
correlator at $p=\om_1$ or at $p=0$.

After evaluating the integrals, we performed the continuum extrapolations
in a similar manner as was discussed in~Sec.~\ref{sec:HE,cont.lim.}.
The continuum limit fit form is
\beq
	f(a/\beta,A_0,{\bf A}^{\delta}) = A_0 + A_1^{\delta} \times\,(a/L_t)^2,
	\label{eq:lin.cont.fit}
\eeq
where $A_0$ is the estimate of the continuum limit from a particular fit,
and the $A_1^{\delta}$ parameters characterize the approach to the continuum 
of the $H_E$ values calculated using the different discretized correlators.
$\delta$ stands for LL, LC or CC.

First, we present results that were obtained using the simple alternative
subtraction of Eq.~(\ref{eq:HEsub,cont,def}) but with $p=\om_1$ or $p=0$.
Although the lattice artefacts are huge in the case of $H_{E,p=0}(\om_2)/T^2$
and $H_{E,p=\om_1}(\om_2)/T^2$, the fit qualities of the continuum extrapolations
using Eq.~(\ref{eq:lin.cont.fit}), i.e.\ a linear fit ansätze in $a^2$ turned out 
to be acceptable.
For instance, only around $\sim 2\%$ of all the $p$-values is smaller than 0.05.
The estimates in the continuum using these subtractions are:
\begin{align}
	\lim_{a^2 \to 0} H_{E,p=0}(\om_2;a)/T^2 &= -0.74(8)_{\rm{stat}}(5)_{\rm{sys}}, \\
	\lim_{a^2 \to 0} H_{E,p=\om_1}(\om_2;a)/T^2 &= -0.71(12)_{\rm{stat}}(4)_{\rm{sys}}.
\end{align}
Depending on the actual modelling interval, the results could be slightly different,
but stay consistent within errors.
These subtractions bring only a modest improvement: the modelling is more precise,
but the slope of the continuum extrapolations are huge (see Fig.~\ref{fig:HEn2,altsubI,contlim}).

Applying the more general subtractions of Eq.~(\ref{eq:HEsub,cont,def,gen})
and Eq.~(\ref{eq:HEn-HEr,cont,def,gen}), we determined $H_E(\om_2)/T^2$
using a broad set of parameters.
We discussed a subset of these results in Secs.~\ref{sec:alter_sub,HE2} and ~\ref{sec:HE2-HE1}.
The results employing a subtraction with vanishing $\varepsilon$ are labelled as fitI
and are listed in Table~\ref{tab:HE2,fitI}.
For those subtractions we used a single parameter, $\alpha$, with which
the integrand can be described as in Eq.~(\ref{eq:altsub,intnd}).
Besides the parameters that characterize the start of the modelling of the
non-static as well as the static screening correlators using single state fits
($T x_{ w,{\rm{ns}} }$ and $T x_{ w,{\rm{st}} }$, respectively), we also list 
the fit parameters ($A_1^{\rm{LL}}, A_1^{\rm{LC}}, A_1^{\rm{CC}}$)
that characterize the slopes of the continuum extrapolations
in Table~\ref{tab:HE2,fitI}.
Using these values, one can read off that the subtraction of the form of
Eq.~\ref{eq:altsub,intnd} gives the flattest continuum extrapolation with
the choice of $\alpha=3.5$.
With this value, the slope parameters of the continuum extrapolation are
consistent with zero for the local-conserved and conserved-conserved 
discretizations.
We summarize the various results for $H_E(\om_2)/T^2$ using these type of
alternative subtractions in Fig.~\ref{fig:HE2,fitI}.

Turning to the subtractions that involve the non-static correlators
at $\om_2$ and $\om_1$ as well, we applied a similar 'two-interval' modelling 
procedure as we discussed above, but used a larger $T x_{ w,{\rm{st}} }$ value for
the non-static screening correlator in the first Matsubara sector as well.
The results employing this type of subtraction with non-vanishing $\varepsilon$ 
are labelled as fitII and are listed in Table~\ref{tab:HE2,fitII}.
In Table~\ref{tab:HE2,fitII}, all results correspond to the choice
$(Tx_{ w,{\rm{ns}} },Tx_{ w,{\rm{st}} })=(0.7,1.1)$; as we did 
earlier (see e.g.\ in Table~\ref{tab:HE2,fitI}), we performed a scan varying
these parameters, but found only small changes.

It is useful to recall that the subtractions using Eq.~(\ref{eq:HEn-HEr,cont,def,gen})
(see~also~Eq.~(\ref{eq:altsub,intnd,a0})) enable a direct calculation 
of the difference of $H_E$s in different Matsubara sectors.
We discussed two of these type of results in Sec.~\ref{sec:HE2-HE1}.
In Table~\ref{tab:HE2,fitII}, more of these types of results are reviewed.
Since the continuum extrapolated quantity is $[H_E(\om_2)-\varepsilon H_E(\om_1)]_{\boldsymbol{p},\boldsymbol{\alpha}}$ 
in this case, we also included a column denoted by $A_0$, that explicitly 
contains this estimate.
By adding $\varepsilon \cdot H_E(\om_1)/T^2$ to this value, one can obtain
an estimate for $H_E(\om_2)/T^2$.
We did this in a correlated way, and then estimated the statistical error on
$H_E(\om_2)/T^2$ and added the systematic errors in quadrature.
The results of Table~\ref{tab:HE2,fitII} are summarized for better overview 
in Fig.~\ref{fig:HE2,fitII}.

\begin{figure}[h]
\begin{center}
\includegraphics[scale=1.00]{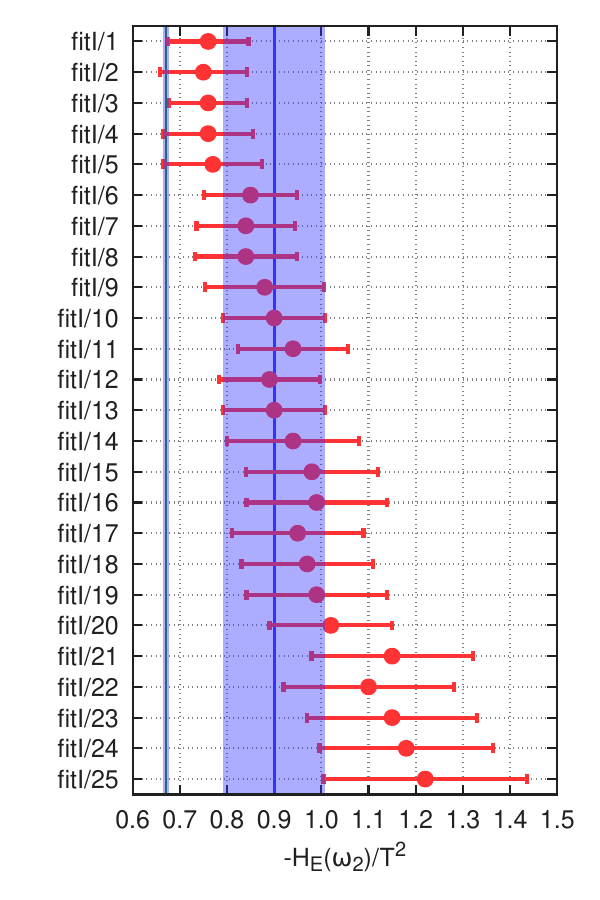}
\caption{
Comparison of continuum extrapolations for $H_E(\om_2)/T^2$ using
various subtractions.
The parameters of the various fits are listed in Table \ref{tab:HE2,fitI}.
The final estimate for $H_E(\om_2)/T^2$ is shown with a blue vertical band.
The final estimate for $H_E(\om_1)/T^2$ is also shown for comparison
with a green vertical band.
}
\label{fig:HE2,fitI}
\end{center}
\end{figure}

\begin{table}[h!]
\begin{tabular}{cccccccc}
\hline
\hline
	fit	&	$\alpha$ & $Tx_{\rm{w,ns}}$ & $Tx_{\rm{w,st}}$ & $-H_E(\om_2)/T^2$ & $A_1^{\rm{LL}}$ & $A_1^{\rm{LC}}$ & $A_1^{\rm{CC}}$ \\ \hline
	fitI/1 & 2.0 & 0.6 & 1.1 & 0.76(8)(3)	& 540(31)(11) & 294(37)(15) & 390(30)(12) \\ \hline
	fitI/2 & 2.0 & 0.7 & 0.9 & 0.75(9)(2)	& 542(36)(10) & 295(30)(16) & 388(30)(11) \\ \hline
	fitI/3 & 2.0 & 0.7 & 1.0 & 0.76(8)(2)	& 540(30)(11) & 295(36)(15) & 390(30)(12) \\ \hline
	fitI/4 & 2.0 & 0.7 & 1.1 & 0.76(9)(3)	& 539(38)(11) & 292(39)(16) & 387(40)(12) \\ \hline
	fitI/5 & 2.0 & 0.7 & 1.3 & 0.77(10)(3) 	& 544(44)(12) & 297(39)(20) & 396(42)(16) \\ \hline
	fitI/6 & 3.0 & 0.6 & 1.1 & 0.85(9)(4)	& 266(48)(14)  	& 131(45)(21)  	& 193(46)(16) \\ \hline
	fitI/7 & 3.0 & 0.7 & 1.0 & 0.84(10)(3) 	& 268(37)(14)  	& 135(49)(18)  	& 193(36)(16) \\ \hline
	fitI/8 & 3.0 & 0.7 & 1.1 & 0.84(10)(4) 	& 265(43)(14)  	& 128(48)(22)  	& 191(40)(16) \\ \hline
	fitI/9 & 3.0 & 0.7 & 1.3 & 0.88(12)(4) 	& 279(56)(16)  	& 144(53)(24)  	& 208(47)(23) \\ \hline
	fitI/10 & 3.5 & 0.6 & 1.1 & 0.90(10)(4) 	& 90(43)(16) 	& 25(51)(27) 	& 65(56)(18) \\ \hline 
	fitI/11 & 3.5 & 0.6 & 1.3 & 0.94(11)(4)	& 101(55)(18) 	& 36(55)(29) 	& 79(56)(19) \\ \hline
	fitI/12 & 3.5 & 0.7 & 1.0 & 0.89(10)(4)	& 88(48)(16) 	& 25(51)(25) 	& 62(44)(18) \\ \hline
	fitI/13 & 3.5 & 0.7 & 1.1 & 0.90(10)(4) 	& 89(49)(17) 	& 24(51)(27) 	& 63(52)(18) \\ \hline
	fitI/14 & 3.5 & 0.7 & 1.3 & 0.94(13)(5) 	& 101(57)(19) 	& 36(51)(31) 	& 81(63)(24) \\ \hline
	fitI/15 & 4.0 & 0.6 & 1.1 & 0.98(13)(5) 	& -118(53)(21)  	& -99(56)(36)  	& -82(57)(22) \\ \hline
	fitI/16 & 4.0 & 0.6 & 1.2 & 0.99(14)(5) 	& -114(60)(23)  	& -98(58)(37)  	& -79(60)(23) \\ \hline
	fitI/17 & 4.0 & 0.7 & 1.0 & 0.95(13)(5) 	& -118(57)(18)  	& -101(48)(32)  & -87(51)(20) \\ \hline
	fitI/18 & 4.0 & 0.7 & 1.1 & 0.97(13)(5) 	& -119(60)(21)  	& -100(60)(37)  & -83(60)(22) \\ \hline
	fitI/19 & 4.0 & 0.7 & 1.2 & 0.99(14)(5) 	& -113(57)(23)  	& -98(54)(38)  	& -80(55)(23) \\ \hline
	fitI/20 & 4.0 & 0.7 & 1.3 & 1.02(12)(6) 	& -97(58)(23)  		& -75(60)(26)  	& -60(58)(26) \\ \hline
	fitI/21 & 5.0 & 0.6 & 1.1 & 1.15(16)(6) 	& -601(68)(30)  	& -388(76)(43)  	& -430(72)(33) \\ \hline
	fitI/22 & 5.0 & 0.7 & 1.0 & 1.10(17)(6) 	& -613(64)(28)  	& -394(82)(46)  	& -443(65)(27) \\ \hline
	fitI/23 & 5.0 & 0.7 & 1.1 & 1.15(17)(6) 	& -601(71)(30)  	& -388(71)(43)  	& -431(73)(33) \\ \hline
	fitI/24 & 5.0 & 0.7 & 1.2 & 1.18(17)(7) 	& -589(85)(31)  	& -379(75)(41)  	& -421(78)(35) \\ \hline
	fitI/25 & 5.0 & 0.7 & 1.3 & 1.22(20)(8) 	& -574(77)(30)  	& -360(75)(46)  	& -403(80)(35) \\ \hline
\hline
\end{tabular}
\caption{
Results for $H_E(\om_2)/T^2$ using alternative subtractions
of the form given in Eq.~(\ref{eq:HEsub,cont,def,gen}).
The values of the momentum in the subtraction given in Eq.~(\ref{eq:HEsub,cont,def,gen})
are $p_1=\om_1$ and $p_2=0$.
$\alpha$ is a single parameter in Eq.~(\ref{eq:HEsub,cont,def,gen}),
that fixes $\alpha_1$ and $\alpha_2$: $\alpha_1=\alpha^2$, $\alpha_2=1-\alpha^2$.
}
\label{tab:HE2,fitI}
\end{table}

\begin{figure}[h]
\begin{center}
	\includegraphics[scale=1.00]{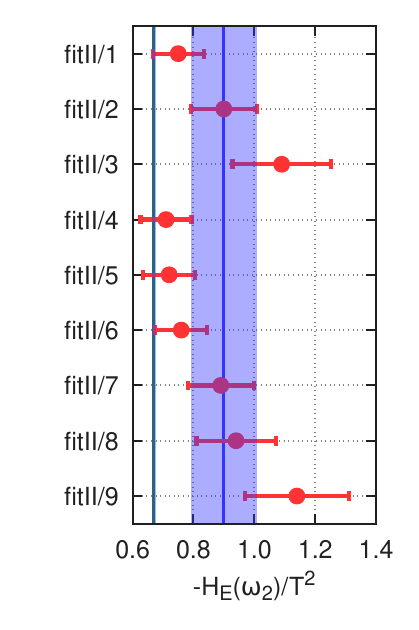}
\caption{
Comparison of continuum extrapolations for $H_E(\om_2)/T^2$ using
various subtractions involving also the non-static correlator at $\om_1$.
The parameters of the various fits are listed in Table \ref{tab:HE2,fitII}.
The final estimate for $H_E(\om_2)/T^2$ is shown with a blue vertical band.
The final estimate for $H_E(\om_1)/T^2$ is also shown for comparison
with a green vertical band.
}
\label{fig:HE2,fitII}
\end{center}
\end{figure}

\begin{table}[h!]
\begin{tabular}{ccccccccc}
\hline
\hline
	fit	&	$\varepsilon$ & $\alpha_1$ & $\alpha_2$ & $-H_E(\om_2)/T^2$ & $A_0$ & $A_1^{\rm{LL}}$ & $A_1^{\rm{LC}}$ & $A_1^{\rm{CC}}$ \\ \hline
fitII/1 & 1     & 3         & -3        & 0.75(8)(3) 	& -0.089(88)(25)     & 544(36)(11)    & 298(34)(15)    & 392(32)(12) \\ \hline
fitII/2 & 1     & 11.25     & -11.25    & 0.90(10)(4) 	& -0.23(12)(4)      & 93(51)(17)     & 31(51)(21)     & 65(47)(18) \\ \hline
fitII/3 & 1     & 24     	& -24	    & 1.09(15)(6) 	& -0.43(17)(6)      & -609(68)(28)     & -391(72)(45)     & -442(66)(26) \\ \hline
fitII/4 & 1     & 0         & 0         & 0.71(8)(3) 	& -0.036(89)(25)     & 707(39)(11)    & 394(36)(15)    & 511(39)(11) \\ \hline
fitII/5 & 2     & 0         & -1        & 0.72(8)(3) 	& 0.62(9)(3)       & 656(35)(11)    & 365(32)(16)    & 474(35)(11) \\ \hline
fitII/6 & 4     & 0         & -3        & 0.76(8)(3) 	& 1.92(8)(3)       & 554(37)(12)    & 306(37)(17)    & 398(40)(12) \\ \hline
fitII/7 & 12.25 & 0         & -11.25    & 0.89(10)(4) 	& 7.30(10)(4)     & 136(42)(17)    & 64(44)(25)     & 86(46)(18) \\ \hline
fitII/8 & 14 	& 0         & -13    	& 0.94(12)(5) 	& 8.44(11)(5)     & 49(40)(18)     & 13(46)(26)     & 20(49)(20) \\ \hline
fitII/9 & 25 	& 0         & -24    	& 1.14(16)(6) 	& 15.62(15)(6)    & -502(53)(26)     & -308(63)(33)     & -396(67)(26) \\ \hline
\end{tabular}
\caption{
Results for $H_E(\om_2)/T^2$ using alternative subtractions with $\varepsilon \ne 0$.
The formula for the integrand is given in Eq.~(\ref{eq:HEsub,cont,def,gen}).
The continuum limit fit ansatz parameters (see Eq.~\ref{eq:lin.cont.fit}) are also listed in the table.
}
\label{tab:HE2,fitII}
\end{table}

\section{Free theory computation of the integrand for $H_E(\om_1)$}
\label{app:free_th}

In this appendix, we provide the expressions for the correlators of interest in the case of non-interacting quarks.
Let $G^{(-1/2)}_{m}(x_3) = e^{-m|x_3|}/(2m)$ be the one-dimensional scalar propagator.
Consider the function
\begin{align}
H(\vec q,\vec s,x_3,\tilde x_3) &= \int \frac{d^{d-1}p}{(2\pi)^{d-1}}\, e^{i\vec p\cdot \vec s}\, 
	G^{(-1/2)}_{\sqrt{\vec p^2+m^2}}(x_3) 
	G^{(-1/2)}_{\sqrt{(\vec p+\vec q)^2+m^2}}(\tilde x_3) \\
	&= \int \frac{d^{d-1}p}{(2\pi)^{d-1}}\; e^{i\vec p\cdot \vec s}
	\frac{1}{4\pi} \int_0^\infty \frac{dt}{\sqrt{t}}\,e^{-t(\vec{p}^{\,\,2}+m^2)-x_3^2/(4t)}
	\int_0^\infty \frac{du}{\sqrt{u}}\,e^{-u((\vec p+\vec q)^2+m^2)-\tilde x_3^2/(4u)} \nonumber \\
	&= \frac{1}{32\pi^{5/2}}\int_0^\infty \frac{dt}{\sqrt{t}}\int_0^\infty \frac{du}{\sqrt{u}}\, \frac{1}{(t+u)^{3/2}} \\
	&\quad \,\exp\Big[-(x_3^2/t+\tilde x_3^2/u)/4 - m^2(t+u)- u \vec{q}^{\,\,2}+(u\vec q-i\vec s/2)^2/(t+u)\Big].
\label{eq:Hwinding}
\end{align}
Now we can write
\begin{align}
G^{\rm T}_{\rm{ns}}(\om_n,x_3) &\equiv -\int \sd x_1 \sd x_2\int_0^\beta \sd x_0 \,\eexp^{i\omega_n x_0} \langle J_1(x) J_1(0) \rangle \\
	&= \frac{N_c}{\beta}\sum_{p_0}^F \int \frac{d^2p_\perp}{(2\pi)^2}
\frac{e^{-(E_{\vec p} + E_{\vec p+\omega_n \vec e_0})|x_3|}}{E_{\vec p} \,E_{\vec p+\omega_n \vec e_0}}
\Big(E_{\vec p}E_{\vec p+\omega_n \vec e_0} + p_0(p_0+\omega_n) + m^2 \Big) \label{eq:Gnsfree_v1} \\
	&= 4N_c \sum_{n_s\in\mathbb{Z}}(-1)^{n_s}\Big(\frac{\partial^2}{\partial x_3\partial \tilde x_3} 
	- \frac{\partial^2}{\partial s_0^2} 
	- i\omega_n\frac{\partial}{\partial s_0} + m^2 \Big)
	H(\vec q,\vec s,x_3,\tilde x_3)_{\tilde x_3=x_3,\vec q = \omega_n \vec e_0,\vec s = 2\pi n_s\beta \vec e_0} 
\end{align}
(where we have included the color factor $N_c=3$ explicitly) and
\begin{align}
G^{\rm T}_{\rm{st}}(\om_n,x_3) &\equiv -\int \sd x_1 \sd x_2 \int_0^\beta \sd x_0 \,\eexp^{i\omega_n x_2} \langle J_1(x) J_1(0) \rangle \\
	&= \frac{N_c}{\beta}\sum_{p_0}^F \int \frac{d^2p_\perp}{(2\pi)^2}
\frac{e^{-(E_{\vec p} + E_{\vec p+\omega_n \vec e_2})|x_3|}}{E_{\vec p} \,E_{\vec p+\omega_n \vec e_2}}
\Big(E_{\vec p}E_{\vec p+\omega_n \vec e_2} -p_1^2 + p_2(p_2+\omega_n) + p_0^2 + m^2 \Big) \label{eq:Gstfree_v1} \\
	&= 4N_c \sum_{n_s\in\mathbb{Z}}(-1)^{n_s}\Big(\frac{\partial^2}{\partial x_3\partial \tilde x_3}
+ \frac{\partial^2}{\partial s_1^2} 
- \frac{\partial^2}{\partial s_2^2}
- i\omega_n\frac{\partial}{\partial s_2}
- \frac{\partial^2}{\partial s_0^2}
+ m^2 \Big) \nonumber \\
	&\quad\, H(\vec q,\vec s,x_3,\tilde x_3)_{\tilde x_3=x_3,\vec q = \omega_n \vec e_2,\vec s = 2\pi n_s\beta\vec e_0}.
\end{align}
The integrand for $H_E(\omega_n,Q^2=0)$, defined to be negative-definite, is given by the difference of the two preceding correlators:
\beq 
H_E(\omega_n,Q^2=0) = 2\int_0^\infty \sd x_3 \,\cosh(\omega_n x_3)\, (G_{\rm ns}^{\rm T}(\om_n,x_3)-G_{\rm st}^{\rm T}(\om_n,x_3)).
\eeq
Using the representation (\ref{eq:Hwinding}) of the function $H$, the `no-winding' term $n_s=0$ cancels in the difference
$G_{\rm ns}^{\rm T}(\om_n,x_3)-G_{\rm st}^{\rm T}(\om_n,x_3)$; this cancellation corresponds to the fact that the integrand
vanishes in the vacuum. The integrand can then be evaluated efficiently at small $x_3$, even directly at $x_3=0$.
At $x_3\gtrsim1/(2\pi T)$ on the other hand, the expressions (\ref{eq:Gnsfree_v1}) and (\ref{eq:Gstfree_v1})
should be used to evaluate the correlator as a rapidly converging sum.

\section{Lattice perturbation theory predictions for $G_E^{\rm T}$
\label{app:lattpt}}

The free Wilson quark propagator is diagonal in color space and can be written in the time-momentum representation as 
\begin{eqnarray}
\langle \psi(x)\;\bar\psi(y)\rangle &\stackrel{x_0\neq y_0}{=} &
\int_{-\pi/a}^{\pi/a} \frac{d^3\vec p}{(2\pi)^3} \frac{e^{-\omega_{\vec p}|x_0-y_0|+i\vec p\cdot(\vec x-\vec y)}}{{\cal D}_{\vec p}} 
\\ && 
\left({\rm sgn}(x_0-y_0) \frac{1}{a}\sinh(a\omega_{\vec p})\gamma_0 -i\vec\gamma\cdot\vec{\circp} 
 + E(x_0-y_0,\vec p)
 \right)
\nonumber
\end{eqnarray}
with the natural convention ${\rm sgn}(0) = 0$ for the sign function and the standard notation
\begin{eqnarray}
\hatp_\mu = \frac{2}{a} \sin\frac{ap_\mu}{2},\qquad \qquad 
\circp_\mu = \frac{1}{a} \sin ap_\mu.
\end{eqnarray}
The constants appearing in the numerator are given by
\begin{eqnarray}
A(\vec p) &=& 1 + am + \frac{1}{2} a^2 \hat{\vec p}^{\,2},
\\
B(\vec p) &=& m^2 + (1+am) \hat{\vec p}^{\,2} + \frac{1}{2} a^2 \sum_{k<l} \hatp_k^2 \hatp_l^2
\end{eqnarray}
The single-quark energy pole is given by
\begin{equation}
\omega_{\vec p} = \frac{2}{a} {\rm \,asinh\,}\left( \frac{a}{2}\sqrt{B(\vec p)/A(\vec p)}\right).
\end{equation}
Further constants are
\begin{eqnarray}
C(\vec p) &=& \frac{1}{2} a\hat{\vec p}^{\,2} + m -\frac{aB(\vec p)}{2A(\vec p)},
\\ 
E(x_0,\vec p) &=& C(\vec p) + \delta_{x_0,0} \frac{\sinh(a\omega_{\vec p})}{a}.
\end{eqnarray}
The denominator reads
\begin{equation}
{\cal D}_{\vec p} = \frac{2}{a}A(\vec p) \sinh(a\omega_{\vec p}) = \sqrt{B(\vec p)\,(4A(\vec p) + a^2 B(\vec p))}.
\end{equation}

After these preliminaries, we are ready to compute the lattice screening correlators at the one loop level.
For that purpose, we use $x_3$ as the `time' direction, so that the quark correlator falls off like $\exp(-\omega_{\vec p}|x_3|)$.
Here we give the result for the local-conserved discretization; the expression of the local and the conserved currents are given in
Eqs.\ (\ref{eq:Vloc,bare}) and (\ref{eq:Vcons,bare}) respectively.
Let $\vec k = (k_0=\omega_n,k_1,k_2)$ be the external momentum, $\omega_n = 2\pi Tn$, $n=0,1,\dots N_t-1$.
Define
\begin{equation}
  G_{11}^{\rm CL}(\vec k; x_3) = - a^3 \sum_{x_0,x_1,x_2} \Big\langle V_1^{\rm C}(x) ~
   V_1^{\rm L}(0)
  \Big\rangle\;
e^{i (\omega_n x_0 + k_1 (x_1+a/2) + k_2 x_2)}.
\end{equation}
With $\vec p = (p_{0},p_1,p_2) = ((2\nu +1)\pi T,p_1,p_2)$, $\nu=0,\dots,N_t-1$, the free-quark prediction is 
\begin{equation}
G_{11}^{\rm CL}(\vec k; x_3) = 4 N_c T \sum_{p_0}^F
\int_{-\pi/a}^{\pi/a} \frac{dp_1\, dp_2}{(2\pi)^2} \;\frac{e^{-(\omega_{\vec p} + \omega_{\vec q})|x_3|}}{{\cal D}_{\vec p}{\cal  D}_{\vec q}}
\;I(x_3;\vec p,\vec q) \Big|_{\vec q = \vec p + \vec k},
\end{equation}
where $N_c=3 $ is the color factor and 
\begin{eqnarray}
I(x_3;\vec p,\vec q) &=&  -\sin\Big( ap_1 + \frac{ak_1}{2}\Big)\bigg( \circp_1 E(x_3,\vec q) + \circq_1 E(x_3,\vec p)\bigg)
\nonumber\\ &&  + \cos\Big(ap_1 + \frac{ak_1}{2}\Big) \bigg( (1-\delta_{x_3,0}) \frac{\sinh(a\omega_{\vec p}) \sinh(a\omega_{\vec q})}{a^2}
\nonumber  \\ && -  \circp_1 \circq_1 + \circp_0 \circq_0 + \circp_2 \circq_2 + E(x_3,\vec p) E(x_3,\vec q)\bigg).
\nonumber
\end{eqnarray}

We are interested in the massless theory, in which case no renormalization factor or additive O($a$)-improvement is needed
at leading order in perturbation theory, since in continuum perturbation theory the vector-tensor correlator vanishes in the chiral limit.
Thus we expect $G_{11}^{\rm CL}(\vec k; x_3)$ to be O($a$)-improved at a fixed $x_3\neq 0$. 
In the free massless theory, the integral over $x_3$ leading to $H_E$ does not necessarily converge at long distances,
due to the single-quark lattice dispersion relation being modified by O$(a^2)$ from its continuum counterpart.
Therefore, in order to judge the size of cutoff effects, we consider in each Matsubara sector the (discretized) truncated integrals
\begin{eqnarray}
{\cal H}_E^{\rm I}(\omega_n) &= & 2 a \sum_{x_3=0}^\beta w(x_3) \,\cosh(\omega_n x_3) G_{\rm ns}(\omega_n,x_3)
\\
{\cal H}_E^{\rm II}(\omega_n) &= & 2 a \sum_{x_3=0}^\beta w(x_3) \,\cosh(\omega_n x_3) (G_{\rm ns}(\omega_n,x_3) - G_{\rm st}(\omega_n,x_3)),
\end{eqnarray}
with
\begin{equation}
w(x_3) = \left\{\begin{array}{l@{~~~}l} 1/2 & x_3=0 ~{\rm or}~x_3=\beta \\ 1 & {\rm else}.   \end{array} \right.
\end{equation}
The expressions above correspond to choosing the trapezoidal rule for the corresponding integrals.
In addition, we consider the estimator
\begin{eqnarray}
  {\cal H}_E^{\rm III}(\alpha,\omega_2) &=&
  2 a \sum_{x_3=0}^\beta w(x_3) \,\Big(\cosh(\omega_2 x_3) G_{\rm ns}(\omega_2,x_3) - \alpha^2 \cosh(\omega_1 x_3)G_{\rm st}(\omega_1,x_3)
\nonumber\\ &&  -(1-\alpha^2) G_{\rm st}(0,x_3) \Big).
\end{eqnarray}
The static contributions appearing in ${\cal H}_E^{\rm II}$ and ${\cal H}_E^{\rm III}$ do not vanish as they would if the integral
extended to $x_3=\infty$. Therefore one should not expect ${\cal H}_E^{\rm I}$, ${\cal H}_E^{\rm II}$ and ${\cal H}_E^{\rm III}$
to have the same continuum limit. However, in the interacting theory, we expect the bulk of the discretization errors to come from
the region $0\leq x_3 \leq \beta$. Each of the three quantities leads to a separate estimator for $H_E$ by extending the integral
to $x_3=\infty$. We therefore investigate the cutoff effects on ${\cal H}_E^{\rm I}$,
${\cal H}_E^{\rm II}$ and ${\cal H}_E^{\rm III}$ in order to assess the relative merits of the corresponding estimators for $H_E$.
The remarks around Eq.\ (\ref{eq:G_smallx3}) 
concerning the behaviour of the integrands in the vicinity of $x_3=0$ for the various estimators 
apply in particular to the free case investigated in this appendix.

\begin{table}
  \begin{tabular}{ccc}
    \hline
    \hline
    $N_t$ &   ${\cal H}_E^{\rm I}(\omega_1)/T^2 $    & ${\cal H}_E^{\rm II}(\omega_1)/T^2$  \\
    \hline
    24 & -1.266  & -1.3237 \\  
    48 & -1.387  & -1.3188 \\
    64 & -1.411  & -1.3181 \\
    96 & -1.430  & -1.3176 \\
    \hline
    \hline
  \end{tabular}
  \caption{\label{tab:latpt_w1} Approach to the continuum of different integrals for the first non-zero Matsubara sector $\omega_1$.}
\end{table}

Table \ref{tab:latpt_w1} compares the approach to the continuum for
the two quantities ${\cal H}_E^{\rm I}$ and ${\cal H}_E^{\rm II}$ in
the Matsubara sector $\omega_1$. Clearly, the approach is much faster for the quantity ${\cal H}_E^{\rm II}$.
It is likely related to the fact that the corresponding estimator for $H_E(\omega_1)$ is free of cutoff effects in the vacuum,
and therefore any cutoff effect must depend mainly on the parameter $aT= 1/N_t$.

Table \ref{tab:latpt_w2} provides a similar comparison in the Matsubara sector $\omega_2$.
As one might expect, cutoff effects are overall larger in this sector.
The quantity ${\cal H}_E^{\rm II}$ approaches its continuum limit much faster than ${\cal H}_E^{\rm I}$,
though clearly, if data were only available up to $N_t=24$, an extrapolation would be needed to reach a precision
of a few percent on the continuum result.
In the lattice QCD simulations, however, the static correlator with spatial momentum equal to $\omega_2$ tends to be noisy.
This motivates us to consider alternative subtraction schemes such as ${\cal H}_E^{\rm III}(\alpha,\omega_2)$.
The choice of $\alpha=2$, while providing a smooth $x_3$ integrand, does not improve the approach to the continuum
as compared to ${\cal H}_E^{\rm I}$. On the other hand, a somewhat larger value of $\alpha$ (for instance 3.5) does lead to
reduced cutoff effects. It is worth pointing out that, in the continuum,
the integrals for ${\cal H}_E^{\rm I,II,III}$ all converge to $-0.5$ when extended to $x_3=\infty$.
The quantity ${\cal H}_E^{\rm III}(3.5,\omega_2)$ is already quite close to that value, indicating that
the integrand in this case must be very suppressed for $x_3>\beta$, which is a further desirable feature.

\begin{table}
  \begin{tabular}{ccccc}
    \hline
    \hline
    $N_t$ &   ${\cal H}_E^{\rm I}(\omega_2)/T^2 $  & ${\cal H}_E^{\rm II}(\omega_2)/T^2$ 
    & ${\cal H}_E^{\rm III}(3.5,\omega_2)/T^2$  & ${\cal H}_E^{\rm III}(2.0,\omega_2)/T^2$ \\
    \hline
    24 & 0.5774 & -1.791 & -0.1685 & 0.3360 \\
    48 & -1.201 & -1.563 & -0.4007 & -0.9372 \\ 
    64 & -1.543 & -1.531 & -0.4443 & -1.182 \\ 
    96 & -1.827 & -1.509 & -0.4831 & -1.386 \\
    \hline
    \hline
    \end{tabular}
\caption{\label{tab:latpt_w2} Approach to the continuum of different integrals for the second non-zero Matsubara sector $\omega_2$.}
\end{table}

\section{Weak-coupling theory results for the screening masses}
\label{app:weak-cth}

In this appendix, we overview our results for the screening
masses obtained using weak-coupling theory.

As it has been derived in Ref.~\cite{Brandt:2014uda}, in order to
calculate the spectrum of non-static screening states in weak-coupling theory,
one has to solve an inhomogeneous Schr\"odinger equation.
Interestingly, apart from different normalizations, this has the
same general form as the one in the LPM resummation of the photon
production rate.
By solving the radial part of the homogeneous Schr\"odinger equation
numerically, we can determine the non-static screening energies by plugging 
in the obtained eigenvalues, $\hat{E}^{(l=1)}$, into 
\beq
	E^{(l=1)} = M_{\rm{cm}} + \frac{g_E^2 C_F}{2\pi} \hat{E}^{(l=1)}.
\eeq
Here, 
\begin{align}
	M_{\rm{cm}} &\equiv 2\pi T n + \frac{m_\infty^2}{ 2 M_{\rm{r}} } \\
	M_{\rm{r}}^{-1} &\equiv \frac{1}{\pi T} \Bigg(\frac{1}{2n-1} +1\Bigg) \\
	C_F &= \frac{N_c^2-1}{2N_c}, \quad N_c=3.
\end{align}
The value of the gauge coupling of the dimensionally reduced effective theory
at two loops is determined to be $g_E^2=g^2 T= 3.2(2)T$~\cite{Brandt:2014uda,Braaten:1995jr,Laine:2005ai}
at $T\sim 250$ MeV.

The numerically determined eigenvalues as well as the corresponding
energies for the LO, NLO and EQCD cases are listed 
in the Tables~\ref{tab:Sch,eigvals,LO},~\ref{tab:Sch,eigvals,NLO} 
and~\ref{tab:Sch,eigvals,EQCD}, respectively.

\begin{table}[h!]
\begin{tabular}{ccc}
\hline
\hline
r & $\hat{E}_r^{(l=1)}$ & $E^{(l=1)}/T$ \\ \hline
0 & 1.57552 & 8.03 \\
1 & 2.16976 & 8.44 \\
2 & 2.54487 & 8.69 \\
\hline
\hline
\end{tabular}
\caption{The non-static screening energies at LO.}
\label{tab:Sch,eigvals,LO}
\end{table}
\begin{table}[h!]
\begin{tabular}{ccc}
\hline
\hline
r & $\hat{E}_r^{(l=1)}$ & $E^{(l=1)}/T$ \\ \hline
0 & 3.33338 & 9.23 \\
1 & 5.68129 & 10.82 \\
2 & 7.63452 & 12.15 \\
\hline
\hline
\end{tabular}
\caption{The non-static screening energies using the NLO potential.}
\label{tab:Sch,eigvals,NLO}
\end{table}
\begin{table}[h!]
\begin{tabular}{ccc}
\hline
\hline
r & $\hat{E}_r^{(l=1)}$ & $E^{(l=1)}/T$ \\ \hline
0 & 4.22873 & 9.83 \\
1 & 7.68636 & 12.18 \\
\hline
\hline
\end{tabular}
\caption{The non-static screening energies using the EQCD potential
determined at $T=400$ MeV~\cite{Panero:2013pla}.}
\label{tab:Sch,eigvals,EQCD}
\end{table}

\section{One-derivative operators having an improved overlap onto the ground state in the transverse non-static sector}
\label{app:op_deriv}

In order to consolidate our extraction of the screening masses from the transverse-channel non-static screening correlators,
Eq.~(\ref{eq:Gns,lat}), we look for alternative operators that may overlap better with the low-energy states.
We therefore consider non-static two-point functions of quark bilinears in the free theory at
infinite spatial volume:
\begin{align}
	C[\Gamma, \mathcal{P}; \tilde{\Gamma}, \tilde{\mathcal{P}}](x_3 - w_3) &= \int_0^\beta\, \sd x_0\,\, \eexp^{\I \om_n (x_0-w_0)} \int\, \sd^2 x_{\perp}\, \left\langle \bar{\psi}(x) \Gamma \mathcal{P}(\nabla_{\bf x}) \psi(x)\,\, \bar{\psi}(w) \tilde{\Gamma} \tilde{\mathcal{P}}(\nabla_{\bf w}) \psi(w) \right\rangle \nonumber \\
	&= \frac{-1}{\beta} \sum_{p_0}^F \int\, \frac{\sd^2 p_{\perp}}{(2\pi)^2}\, \frac{ \eexp^{-(E_{\bf p} + E_{{\bf p}+\om_n \hat{e}_0})|x_3-w_3|} }{E_{\bf p}E_{{\bf p} + \om_n \hat{e}_0}}\, \mathcal{P}(\I {\bf p}) \tilde{\mathcal{P}}(\I {\bf p})\, T(\Gamma, \tilde{\Gamma})
\end{align}
where $\Gamma$ is for the Dirac structure, $\mathcal{P}(\nabla)$ is a polynomial
in $\nabla=(\partial/\partial x_1, \partial/\partial x_2)$ 
and ${\rm T}(\Gamma, \tilde{\Gamma})$  denotes the trace:
\beq
	{\rm T}(\Gamma, \tilde{\Gamma}) = \frac{1}{4} {\rm Tr}\left\{ \Gamma\, (E_{\bf p} s_0 \gamma_0 - \I {\bf p}\cdot {\bf \gamma} + m)\, \tilde{\Gamma}\, (-E_{{\bf p}+\om_n \hat{e}_0} s_0 \gamma_0 - \I ({\bf p} + \om_n \hat{e}_0)\cdot {\bf \gamma} + m ) \right\}
\eeq
with $s_0 = \sgn(x_0 - w_0)$.
Time is in the direction ''0''.

For the vector current: $\Gamma = \tilde{\Gamma} = \gamma_1$ 
and $\mathcal{P} = \tilde{\mathcal{P}} = 1$, which results in
\beq
	T(\gamma_1, \gamma_1) = E_{\bf p} E_{{\bf p} + \om_n \hat{e}_0} + p_0 (\om_n + p_0) - p_1^2 + p_2^2 + m^2.
\eeq
Consider the case of the first Matsubara sector, $\om_n=2 \pi T$. The lowest exponential is then realized for $p_3=- \pi T$.
Upon integration, the $(-p_1^2+p_2^2)$ terms cancel in $T(\gamma_1, \gamma_1)$,
due to the O(2) symmetry in the xy plane.
Furthermore, we see that $T(\gamma_1, \gamma_1)$ vanishes at ${\bf p}_{\perp}=0$
in the chiral limit.
The first contribution comes from terms of order $T(\gamma_1, \gamma_1) \sim {\bf p}_{\perp}^2$,
which suppresses the correlator at long distances\footnote{The suppression of the transverse-channel non-static correlator
was already pointed out in~\cite{Brandt:2014uda}.}.
Therefore we seek for such alternative operators that may overlap better 
with low-energy states and have a non-vanishing correlation with $\bar{\psi}\gamma_1 \psi$ even at $m=0$.
We consider operators with one derivative.

There are four of such one-derivative operators,
\beq
	\bar{\psi} \gamma_0 \gamma_5 (\overset{\rightarrow}{D_2} - \overset{\leftarrow}{D_2}) \psi, \qquad \bar{\psi} \gamma_3 \gamma_5 (\overset{\rightarrow}{D_2} - \overset{\leftarrow}{D_2}) \psi
	\label{eq:D,Codd}
\eeq
and
\beq
	\bar{\psi} \gamma_0 (\overset{\rightarrow}{D_1} - \overset{\leftarrow}{D_1}) \psi, \qquad \bar{\psi} \gamma_3 (\overset{\rightarrow}{D_1} - \overset{\leftarrow}{D_1}) \psi
	\label{eq:D,Ceven}
\eeq
which give a non-vanishing Dirac trace when correlated with the 
current $\bar{\psi}\gamma_1 \psi$.

However, when considering the $C$-parity of these operators, one finds that
the ones in Eq.~(\ref{eq:D,Ceven}) are $C$-even.
Recalling that the conserved vector current is $C$-odd, we find that correlation
function of the operators in Eq.~(\ref{eq:D,Ceven}) with $\bar{\psi}\gamma_1 \psi$
vanishes.
Therefore only the operators of Eq.~(\ref{eq:D,Codd}) need to be considered.
The Dirac traces $T(\Gamma,\tilde{\Gamma})$ of these operators
\beq
	T(\Gamma,\tilde{\Gamma}) = {\bf p}_\perp^2 - m^2 \pm (E_{\bf p} E_{{\bf p} + \om_n \hat{e}_0} - p_0 (p_0+\om_n) ),
\eeq
with plus (minus) sign for the operator with $\Gamma=\tilde{\Gamma}= \gamma_0 \gamma_5$ ($\Gamma=\tilde{\Gamma}= \gamma_3 \gamma_5$).
Thus, we find that in the free theory both operators of  Eq.~(\ref{eq:D,Codd}) have an unsuppressed coupling
to low-lying states.

We have investigated the two-point function of the 
second operator of Eq.~(\ref{eq:D,Codd}), now in the interacting theory on our coarsest
ensemble called F7. 
This $12\times 48^3$ ensemble at the same temperature of $T=254$\,MeV has not been included in the main analysis.
Its relevant parameters are given in Table I of Ref.~\cite{Ce:2022fot}.

We introduce the following notation for the lattice operator, 
\beq
	O_{\rm{35D}}(x) = (\bar{\psi} \gamma_3 \gamma_5 \overset{\leftrightarrow}{D_2} \psi)(x)
\eeq
where the forward-backward differential
\beq
	(\bar{\chi} \overset{\leftrightarrow}{D_{\mu}} \psi)(x) = \bar{\chi}(x)(D_\mu \psi)(x) - (D_\mu \bar{\chi})(x) \psi(x)
\eeq
is defined in terms of the symmetric, covariant finite-difference operators
\begin{align}
	D_\mu\psi(x) &= \frac{1}{2a} \left[ U_\mu(x) \psi(x+a\hat{\mu}) - U_\mu^\dagger(x-a\hat{\mu}) \psi(x-a\hat{\mu}) \right] \\
	D_\mu\bar{\psi}(x) &= \frac{1}{2a} \left[ \bar{\psi}(x+a\hat{\mu}) U_\mu^\dagger(x) - \bar{\psi}(x-a\hat{\mu}) U_\mu(x-a\hat{\mu}) \right].
\end{align}

\begin{figure}[t]
\begin{center}
\includegraphics[scale=0.80]{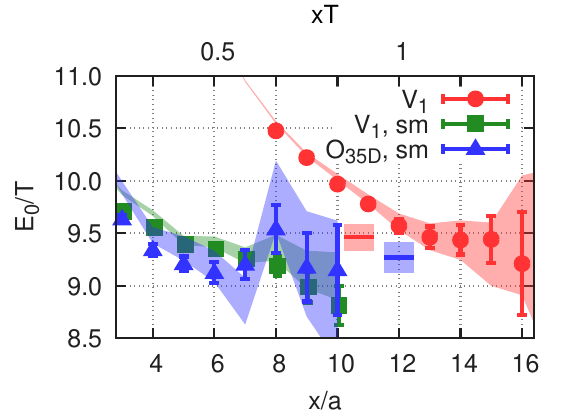}
\caption{
Fitted and local effective (cosh) masses with symbols and bands, respectively,
extracted from the two-point correlation functions of the $V_1$, smeared $V_1$
and smeared $O_{\rm{35D}}$ operators with red circles, green squares and blue triangles, 
respectively.
The short, horizontal red and blue bands show the value and error of the window-averaged masses.
}
\label{fig:F7,wmass}
\end{center}
\end{figure}

Using the above definitions, we then calculated the two-point correlation function 
of the $O_{\rm{35D}}(x)$ operator.
We performed this evaluation using also a smeared operator, which
reduced the noise significantly.
We also note that, similarly to the transverse screening correlators of the vector current,
we performed averages over different spatial decay directions
exploiting the discrete lattice rotation symmetry.

We extracted the local effective mass from the two-point function
and we also performed single-state fits using a similar fit ansatz
as in Eq.~(\ref{eq:ns,scrcorr,asympt}).
Using the fit results for the mass, we constructed the window-smeared
mass averaging fit results obtained at subsequent equal-length fit ranges.
The window function was centered at $xT=0.6$ and had a width of 0.30 in this variable.
These correspond to $x_{\rm{win,center}}/a \approx 7$ and a width of $3.6$ in
lattice units.
We note that  for instance the fit results using 9 datapoints 
starting with $x/a=4,5,6,7,8,9,10$ have $p$-values 
0.02, 0.27, 0.15, 0.18, 0.45, 0.68, 0.59, respectively,
i.e.\ the chosen window range covers results with acceptable $p$-values.
The window-averaged mass is shown by the horizontal blue bland in Fig.~\ref{fig:F7,wmass}.
Compared to this we also show in Fig.~\ref{fig:F7,wmass} the single-state
fit results for the two-point function involving the smeared or unsmeared
$V_1$ operator.
The window-smeared mass obtained using the unsmeared $V_1$ is shown with
the short red horizontal band in Fig.~\ref{fig:F7,wmass} and is in good
agreement with the one obtained using $O_{\rm{35D}}(x)$.

In summary, while the smeared $O_{\rm{35D}}$ and $V_1$ correlators
exhibit lower effective masses in the region $x_3T\lesssim0.75$, their
variance is also larger. This observation should be taken into account
in future dedicated spectroscopic calculations.


\bibliography{imamom_n1}

\end{document}